\documentclass[aps,prd,twocolumn,showpacs,floats,floatfix,letterpaper,nofootinbib,superscriptaddress,]{revtex4-1}

\usepackage{hyperref}
\usepackage{amssymb,amsmath,latexsym,mathrsfs}
\usepackage{graphicx,subfigure}
\usepackage{epsfig}
\usepackage{varioref,xr-hyper}
\usepackage{color}
\usepackage{multirow}
\usepackage{array}
\hypersetup{
     colorlinks   = true,
     citecolor    = blue
     }

\begin{document}

\title{Methods for studying the accuracy of light propagation in N-body simulations}
\author{S. M. Koksbang}
\email{koksbang@phys.au.dk}
\affiliation{Department of Physics and Astronomy, Aarhus University, 8000 Aarhus C, Denmark}
\author{S. Hannestad}
\affiliation{Department of Physics and Astronomy, Aarhus University, 8000 Aarhus C, Denmark}
\affiliation{Aarhus Institute of Advanced Studies, Aarhus University, DK-8000 Aarhus C, Denmark}

\begin{abstract}
It is proposed to use exact, cosmologically relevant solutions to Einstein's equations to accurately quantify the precision of ray tracing techniques through Newtonian N-body simulations. As an initial example of such a study, the recipe in (Green $\&$ Wald, 2012) for going between N-body results and a perturbed FLRW metric in the Newtonian gauge is used to study light propagation through quasi-spherical Szekeres models. The study is conducted by deriving a set of ODEs giving an expression for the angular diameter distance in the Newtonian gauge metric. The accuracy of the results obtained from the ODEs is estimated by using the ODEs to determine the distance-redshift relation in mock N-body data based on quasi-spherical Szekeres models. The results are then compared to the exact relations. From this comparison it is seen that the obtained ODEs can accurately reproduce the distance-redshift relation along both radial and non-radial geodesics in spherically symmetric models. The reproduction of geodesics in non-symmetric Szekeres models is slightly less accurate, but still good. These results indicate that the employment of perturbed FLRW metrics for standard ray tracing techniques yields fairly accurate results, at least regarding distance-redshift relations. It is possible though, that this conclusion will be rendered invalid if other typical ray tracing approximations are included and if light is allowed to travel through several structures instead of just one.
\end{abstract}


\maketitle

\section{Introduction}
Supernovae observations are typically interpreted as indicating an accelerated expansion of the Universe \cite{supernova1, supernova2}. Such interpretations, leading to the inclusion of a cosmological constant into the cosmological standard model, are predominantly based on redshift-distance relations valid only within the Friedmann-Lemaitre-Robertson-Walker (FLRW) models \cite{Durrer}. Since the Universe is not exactly homogeneous and isotropic, it is important to know how deviations from an exact FLRW universe affect light propagation. By using exact inhomogeneous solutions to Einstein's equation it has {\em e.g} been shown that inhomogeneities can affect light propagation such that it may look as though the Universe is undergoing an accelerated expansion even though it is not (see {\em e.g.} \cite{no_need_for_lambda1,no_need_for_lambda2,no_need_for_lambda3}). As of yet, no convincing results have been obtained indicating that this is in fact what happens in the real universe though. On the contrary actually, studies indicate that randomization and statistical averaging diminishes any deviations from FLRW results \cite{statistics,dallas_cheese}
\footnote{Note added after publication: it has later been pointed out to us, that though this seems to be the case in general when cosmic backreaction is vanishing (see {\em e.g.} \cite{added_later1, added_later2}), it is not necessarily the case when the models studied have non-vanishing backreaction (see {\em e.g.} \cite{added_later3}).}. 
\newline\indent
Even if inhomogeneities cannot explain the seeming accelerated expansion of the Universe, being able to quantify even small effects may be important for parameter determinations based on observations made in an era of precision cosmology.
\newline\newline
Newtonian N-body simulations are an important tool for studying structure formation and are widely accepted as giving a correct description of the structure formation going on in the real universe. Comparing results from N-body simulations with observations is thus essential for using survey data to put restrictions on cosmological parameters determining the standard model of cosmology. Unfortunately, Newtonian mechanics cannot be used to study exact general relativistic light propagation since this requires knowledge of the exact metric. Typically, light propagation through an N-body simulation is thus studied through ray tracing techniques (see {\em e.g.} \cite{arbitrary, raytrace_beskrivelse, bert,ray_trace_pert}), where bundles of light are propagated through a simulation box via the background metric. If at all, only at a discrete number of lens planes is the light deflected based on linearized relativistic gravity. Several improvements of the basic ray tracing scheme have been developed and the accuracy of the methods have been studied through different approximation schemes (see {\em e.g.} \cite{relativistic_Nbody,raytrace_beskrivelse, rayTrace_igen, 3dRayTrace}). For example, in \cite{3dRayTrace} the standard ray-bundle method (see {\em e.g.} \cite{RBM}) was sought improved by using the first order metric, rather than the background metric, to describe null-geodesics. 
\newline\indent
In an era of precision cosmology, it is pertinent to know exactly to what extent theoretical predictions of light propagation through an N-body simulation can be trusted and the work presented here is another step towards that goal.
\newline\newline
In \cite{recipe} it is argued that the perturbed metric in the longitudinal gauge gives a good description of the metric of the Universe. A dictionary giving a recipe for going between Newtonian N-body results and this metric is also given. The work presented here is based on that recipe. Several versions of the recipe are given in \cite{recipe}, varying from the most simple form to a very involved "Oxford" version. The version which in \cite{recipe} is denoted the abridged version is the one which is used here because it is that version which is most compatible with standard ray tracing techniques - and it is the simplest of the recipe versions to use with N-body simulation data. The metric is then simply the usual first order metric in the Newtonian gauge with the potential and velocity fields being those obtained directly from an N-body simulation.
\newline\indent
The compatibility between a first order perturbed metric and the metric of the real universe is not clear. It is advocated in {\em e.g.} \cite{recipe} that it can be used, while others show more skepticism and {\em e.g.} point out dangers with using perturbation theory to justify itself (see {\em e.g} \cite{Syksy_kritik}). Since the perturbed FLRW metric in the Newtonian gauge combined with versions of the recipe of \cite{recipe} are typically the basis of ray tracing through N-body simulations, it is important to study how accurate the use of this recipe really is. In order to do this, the metric and dictionary described above will be used to derive a set of ODEs which can be solved to obtain distance-redshift relations in the corresponding spacetime. To study the accuracy of the obtained relation, the ODEs are used on mock N-body data constructed from quasi-spherical Szekeres models \cite{Szekeres} under the standard assumption that Newtonian N-body simulations accurately reproduce relativistic structure formation. Since the quasi-spherical Szekeres models are exact solutions to Einstein's field equations, the exact distance-redshift relations of these models can be obtained and compared with those obtained by using the recipe.
\newline\indent
There are several reasons for choosing the quasi-spherical Szekeres models for the accuracy checking of the ODEs. First of all, the quasi-spherical Szekeres models are counted amongst the most realistic cosmologically relevant exact known solutions to Einstein's equations. Further more, the spherically symmetric limit of the quasi-spherical Szekeres models, the Lemaitre-Tolman-Bondi (LTB) models, are known to be reproducible by Newtonian N-body simulations \cite{Troels}. Thus, checking the ODEs with LTB models gives both a check of reproducing light propagation through exact non-linear relativistically evolving structures and through structures evolving according to Newtonian N-body simulations. It is not surprising that LTB models are reproducible by Newtonian mechanics since they are spherically symmetric and their structure formation is scale invariant. It is to obtain slightly more generality that also the non-symmetric quasi-spherical Szekeres models are used in the comparison. The non-symmetric Szekeres models show significantly more growth of structure than LTB models \cite{bolejko1, bolejko2,dallas_growth1,dallas_growth2} and whether or not these can be reproduced by N-body simulations is still unknown but is the subject of ongoing work of one of the current authors.
\newline\indent
It should be noted that the perturbed FLRW metric is diagonal while the Szekeres metric is non-diagonal in spherical coordinates. This immediately implies a shortcoming of the Newtonian gauge metric. However, there is no reason to expect the local metric of cluster-void structures in the real universe to be diagonal. Hence, the non-diagonal metric of the Szekeres model may actually be considered a strength for the purpose of the study.

\section{The angular diameter distance in the Newtonian gauge}\label{section:DA_newtonian}
In this section, the ODEs needed to obtain the angular diameter distance $D_A(z)$ in the Newtonian gauge are given. The ODEs are obtained following the procedure presented in \cite{dallas-light}.
\newline\newline
The onset of deriving the sought equations is the metric. Since Szekeres models are dust models, the anisotropic stress vanishes and the perturbed FLRW metric in the Newtonian gauge can be written using a single perturbation field $\psi$:

\begin{equation}\label{eq:lineelement_ng}
\begin{split}
ds^2 = -c^2(1+2\psi)dt^2 + a^2(1-2\psi)\left(dr^2 + r^2d\theta^2 + \sin^2(\theta)d\phi^2 \right)  \\
 = -Tdt^2 + Rdr^2 + Fd\theta^2 + Pd\phi^2
\end{split}
\end{equation}

The metric is assumed to be exactly described in this form and will be referred to as the Newtonian gauge metric. It should be stressed, that though the metric above looks like a perturbed FLRW metric in the Newtonian gauge, there is a small difference: the above metric is an attempt to obtain a relativistic description of the "underlying" relativistic spacetime corresponding to a Newtonian N-body simulation. As such, according to the recipe of \cite{recipe}, the potential $\psi$ is to be obtained from the exact, non-linear density field from the corresponding N-body simulation, and not through perturbation theory. This point will be revisited in section \ref{section_mapping}.
\newline\indent
As implied in the expression for the line element above, spherical coordinates have been used for this work. By writing the metric functions as $T,R,F,P$, the equations derived below are written in a generic form so that they are valid with any coordinate choice and actually any spacetime as long as the metric is diagonal. In appendix \ref{gauge-appendix_sph}, the equations are written in their full length in spherical coordinates using $\psi$ instead of $T,R,F,P$.
\newline\newline
Light moves along null-geodesics, so the geodesic equations are needed to describe the light paths. Letting a dot denote differentiation with respect to the affine parameter $\lambda$, the geodesic equations can be written as:

\begin{equation}\label{eq:kt_T}
-2T\dot k^t = 2\dot T k^t -T_{,t}(k^t)^2 + R_{,t}(k^r)^2 + F_{,t}(k^{\theta})^2 + P_{,t}(k^{\phi})^2 
\end{equation}
\begin{equation}\label{eq:kr_R}
2R\dot k^r = -2\dot R k^r - T_{,r}(k^t)^2 + R_{,r}(k^r)^2 + F_{,r}(k^{\theta})^2 + P_{,r}(k^{\phi})^2
\end{equation}
\begin{equation}
2F\dot k^{\theta} = -2\dot F k^{\theta} - T_{,\theta}(k^t)^2 + R_{,\theta} (k^r)^2 + F_{,\theta}(k^{\theta})^2 + P_{,\theta}(k^{\phi})^2
\end{equation}
\begin{equation}
2P\dot k^{\phi} = - 2\dot P k^{\phi} - T_{,\phi}(k^t)^2 + R_{,\phi}(k^r)^2 + F_{,\phi}(k^{\theta})^2 + P_{,\phi}(k^{\phi})^2
\end{equation}
Subscripted commas followed by coordinates denote partial derivatives with respect to those coordinates.
\newline\newline
Considering a light bundle with infinitesimal cross section $\delta S$ and solid angle element $\delta\Omega$, an ODE for $D_A$ can be obtained as follows (see {\em e.g.} \cite{Sachs}):

\begin{equation}\label{eq:DAalpha}
\delta S = D_A^2\delta\Omega \implies 2d\ln D_{A} = d\ln\delta S = 2\tilde\theta d\lambda = k^{\alpha}_{;\alpha}d\lambda
\end{equation}
$\tilde\theta$ denotes the optical expansion scalar (see {\em e.g.} \cite{Sachs}) and a subscripted semi-colon followed by a coordinate denotes covariant differentiation with respect to that coordinate. Summation over repeated indexes is implied, with Greek indexes running over $0-3$ and Latin indexes running over $1-3$.
\newline\indent
Inserting the appropriate Christoffel symbols (see appendix \ref{gauge-appendix_sph}) into this equation, the following ODE is obtained:
\begin{equation}\label{DA}
\frac{dD_A}{d\lambda} = \frac{1}{2}D_A( k^{\alpha}_{,\alpha} + \frac{\dot T}{T} + \frac{\dot R}{R} + \frac{\dot F}{F} + \frac{\dot P}{P})
\end{equation}
Once this expression has been used to obtain $D_A$ along a geodesic, the luminosity distance can be found as $D_L = (1+z)^2D_A$ and from this the apparent magnitude etc. can be obtained.
\newline\indent From equation (\ref{DA}) it is apparent that $k^t_{,t}, k^r_{,r}, k^{\theta}_{,\theta}$ and $k^{\phi}_{,\phi}$ are needed and these are obtained by solving the ODEs that appear when taking the partial derivatives of the geodesic equations:

\begin{equation}
\begin{split}
-2T\frac{d}{d\lambda}k^t_{,\alpha} = 2T_{,\alpha}\dot k^t + 2Tk^t_{,\beta}k^{\beta}_{,\alpha} + 2k^t(T_{,\alpha\beta}k^{\beta} + T_{,\beta}k^{\beta}_{,\alpha}) +\\
2\dot k^t_{,\alpha} - T_{,t\alpha}(k^t)^2 -
2T_{,t} k^tk^t_{,\alpha} + R_{,t\alpha}(k^r)^2 + 2R_{,t}k^rk^r_{,\alpha} +\\ 
F_{,t\alpha}(k^{\theta})^2 + 2F_{,t}k^{\theta}k^{\theta}_{,\alpha}  +P_{,t\alpha}(k^{\phi})^2 + 2P_{,t}k^{\phi}k^{\phi}_{,\alpha}
\end{split}
\end{equation}
\begin{equation}
\begin{split}
2R\frac{d}{d\lambda}k^r_{,\alpha} = -2R_{,\alpha}\dot k^r - 2Rk^r_{,\beta}k^{\beta}_{,\alpha} - 2k^r(R_{,\alpha\beta}k^{\beta}+ R_{,\alpha}k^{\beta}_{,\alpha})-\\ 2\dot Rk^r_{,\alpha} - T_{,r\alpha}(k^t)^2 -
2T_{,r}k^tk^t_{,\alpha} + R_{,r\alpha}(k^r)^2 + 2R_{,r}k^rk^r_{,\alpha} +\\ 
F_{,r\alpha}(k^{\theta})^2 + 2F_{,r}k^{\theta}k^{\theta}_{,\alpha} + P_{,r\alpha} (k^{\phi})^2 + 2P_{,r}k^{\phi}k^{\phi}_{,\alpha}
\end{split}
\end{equation}
\begin{equation}
\begin{split}
2F\frac{d}{d\lambda}k^{\theta}_{,\alpha} = -2F_{,\alpha}\dot k^{\theta} - 2Fk^{\theta}_{,\beta}k^{\beta}_{,\alpha} - 2k^{\theta}(F_{,\beta\alpha}k^{\beta} + F_{,\beta}k^{\beta}_{,\alpha}) -\\
2\dot F k^{\theta}_{,\alpha} -T_{,\theta\alpha}(k^t)^2 - 2T_{,\theta}k^tk^t_{,\alpha} + R_{,\theta\alpha} (k^r)^2 + 2R_{,\theta}k^rk^r_{,\alpha} +\\ 
F_{,\theta\alpha}(k^{\theta})^2 + 2F_{,\theta}k^{\theta}k^{\theta}_{,\alpha} + P_{,\theta\alpha}(k^{\phi})^2 + 2P_{,\theta}k^{\phi}k^{\phi}_{,\alpha}
\end{split}
\end{equation}
\begin{equation}
\begin{split}
2P\frac{d}{d\lambda}k^{\phi}_{,\alpha} = -2P_{,\alpha}\dot k^{\phi} - 2Pk^{\phi}_{,\beta}k^{\beta}_{,\alpha} - 2k^{\phi}(P_{,\alpha\beta}k^{\beta} + P_{,\beta}k^{\beta}_{,\alpha})\\
-2\dot Pk^{\phi}_{,\alpha} - T_{,\phi\alpha}(k^t)^2 -
2T_{,\phi}k^tk^t_{,\alpha} + R_{,\phi\alpha}(k^r)^2 + 2R_{,\phi}k^rk^r_{,\alpha} +\\ 
F_{,\phi\alpha}(k^{\theta})^2 + 2F_{,\phi}k^{\theta}k^{\theta}_{,\alpha} + P_{,\phi\alpha}(k^{\phi})^2 + 2P_{,\phi}k^{\phi}k^{\phi}_{,\alpha}
\end{split}
\end{equation}
\newline\newline
In order to ensure that the geodesics described by the above equations are null-geodesics, the null condition and its partial derivatives shown below can be used to set the initial conditions:

\begin{equation}
k^{\alpha}k_{\alpha} = -T(k^t)^2 + R(k^r)^2 + F(k^\theta)^2 + P(k^{\phi})^2 = 0
\end{equation}
\begin{equation}
\begin{split}
-T_{,\alpha}(k^t)^2 - 2Tk^tk^t_{,\alpha} + R_{,\alpha}(k^r)^2 + 2Rk^rk^r_{,\alpha} +\\
F_{,\alpha}(k^{\theta})^2 + 2Fk^{\theta}k^{\theta}_{,\alpha} + P_{,\alpha}(k^{\phi})^2 + 2Pk^{\phi}k^{\phi}_{,\alpha} = 0
\end{split}
\end{equation}
\newline\newline
The objective of this work is to asses the accuracy of the equations presented above when applied to N-body data. In order to do this, the exact relativistic spacetime corresponding to the N-body data is needed. The study is thus conducted by constructing mock N-body data corresponding to a quasi-spherical Szekeres model and its underlying LTB model. The above equations are solved with the potentials corresponding to these two "data sets". The results are then compared to those obtained by solving the equivalent ODEs derived from the actual metrics of these models.

\section{Quasi-spherical Szekeres models}
In this section, the quasi-spherical Szekeres models are introduced and the ODEs needed to obtain $D_A(z)$ in these models are given. At the end of the section, the relation between the Szekeres models and the Newtonian gauge metric is discussed.
\newline\newline 
The Szekeres models \cite{Szekeres} is a family of exact, inhomogeneous dust solutions to the Einstein equations which in their general form have no killing vectors \cite{killing}. A typical coordinate system to describe the Szekeres metric in, is the $(t,r,p,q)$-system related to the spherical coordinate system by a stereographic projection \cite{hellaby} \footnote{For notational convenience, the coordinates $t,r,\theta$ and $\phi$ of the Szekeres metric will notationally not be distinguished from the coordinates of the Newtonian gauge metric. The coordinates of these two spacetimes are {\em not} the same though and mappings between the coordinate systems are discussed in section \ref{section_mapping}.}:

\begin{equation}\label{stereo}
\begin{split}
p-\tilde P = S\cot(\theta/2)\cos(\phi)\\
q-Q = S\cot(\theta/2)\sin(\phi)\\
\end{split}
\end{equation}

The functions $\tilde P, Q$ and $S$ are defined below.
\newline\newline
The $(t,r,p,q)$ coordinate system is particularly useful since it renders the metric diagonal:

\begin{equation}\label{eq:line_elemet}
\begin{split}
ds^{2} = -c^2dt^{2} +\frac{\left(A_{,r}(t,r)-A(t,r)\frac{E_{,r}(r,p,q)}{E(r,p,q)}\right)^2}{\epsilon-k(r)}dr^2 +\\
\frac{A(t,r)^2}{E(r,p,q)^2}(dp^2+dq^2)
\end{split}
\end{equation}

In a general Szekeres model, $E$ is given as $E = \frac{1}{2S}(p^2+q^2)-\frac{p\tilde P}{S} - \frac{qQ}{S}+\frac{\tilde P^2+Q^2+\epsilon S^2}{2S}$, where $S, \tilde P$ and $Q$ are continuous but otherwise arbitrary functions of $r$, and $\epsilon \in \{-1,0,1\}$. The quasi-spherical Szekeres models with $\epsilon = 1$ reduce to LTB models when $\tilde P, Q$ and $S$ are constant functions. Only the quasi-spherical Szekeres models are considered in the following. 
\newline \newline
Inserting the metric corresponding to the line element (\ref{eq:line_elemet}) into Einstein's equation for a dust-filled universe containing a cosmological constant leads to the following two equations:
\begin{equation}\label{eq:friedmann}
\frac{1}{c^2}A_{,t}^2 = \frac{2M}{A}-k+\frac{1}{3c^2}\Lambda A^2
\end{equation}
\begin{equation}\label{eq:density}
\rho = \frac{2M_{,r}-6M\frac{E_{,r}}{E}}{c^2\beta A^2(A_{,r}-A\frac{E_{,r}}{E})},\, \, \, \, \, \, \, \, \beta = 8\pi G/c^4
\end{equation}
The function $M = M(r)$ appearing in these equations is a temporal integration constant depending on the radial coordinate and corresponds to the effective gravitational mass at comoving radial coordinate $r$.

\subsection{Model setup}\label{models}
\begin{figure}
\centerline{\includegraphics[scale = 0.5]{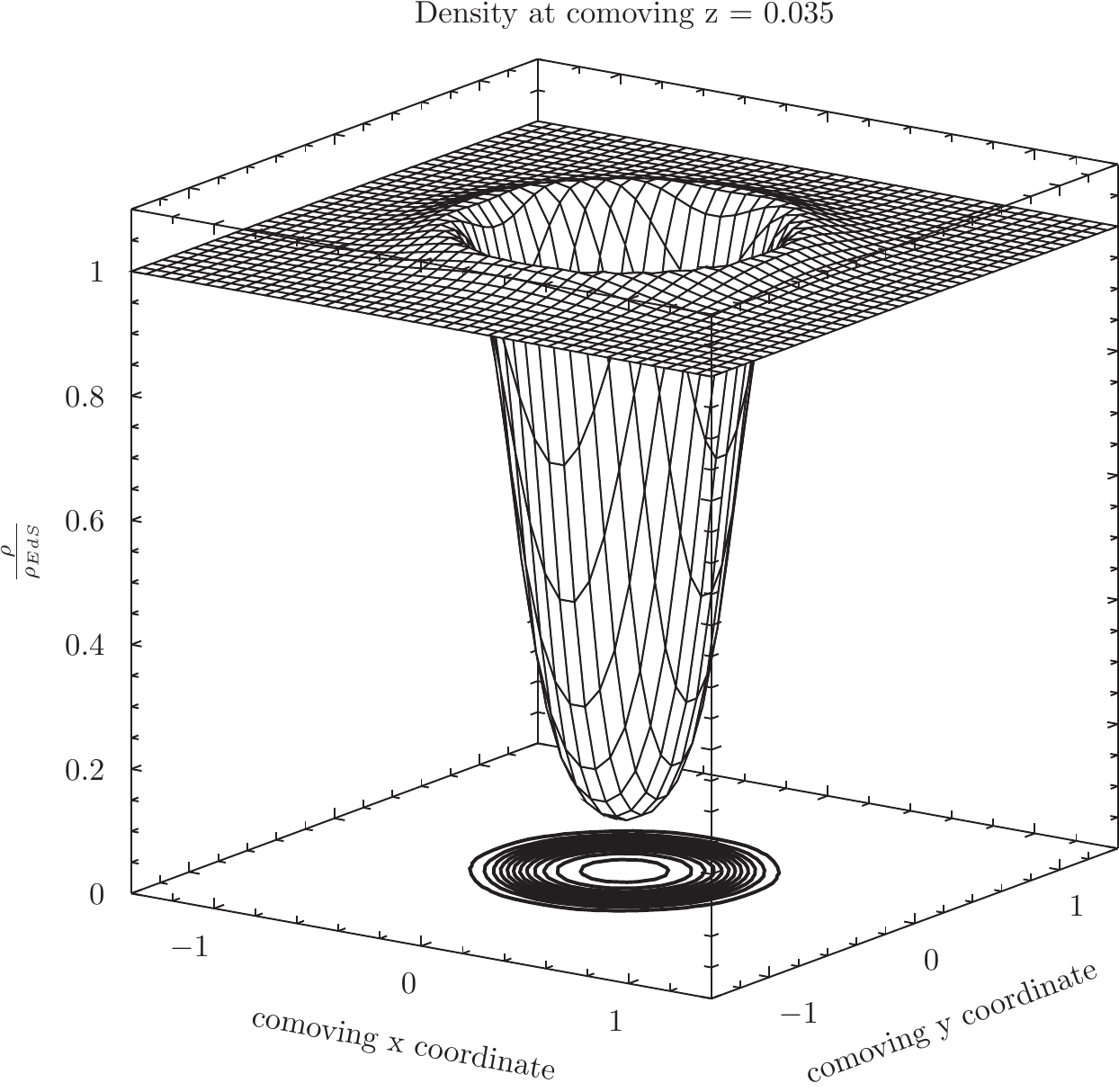}}
\caption{Present time 2D density profile of the LTB model. The comoving coordinates are normalized at $t = t_{ls}$ in units of $0.1Mpc$.}
\label{ltb_density_2d}
\end{figure}

\begin{figure}
\centerline{\includegraphics[scale = 0.5]{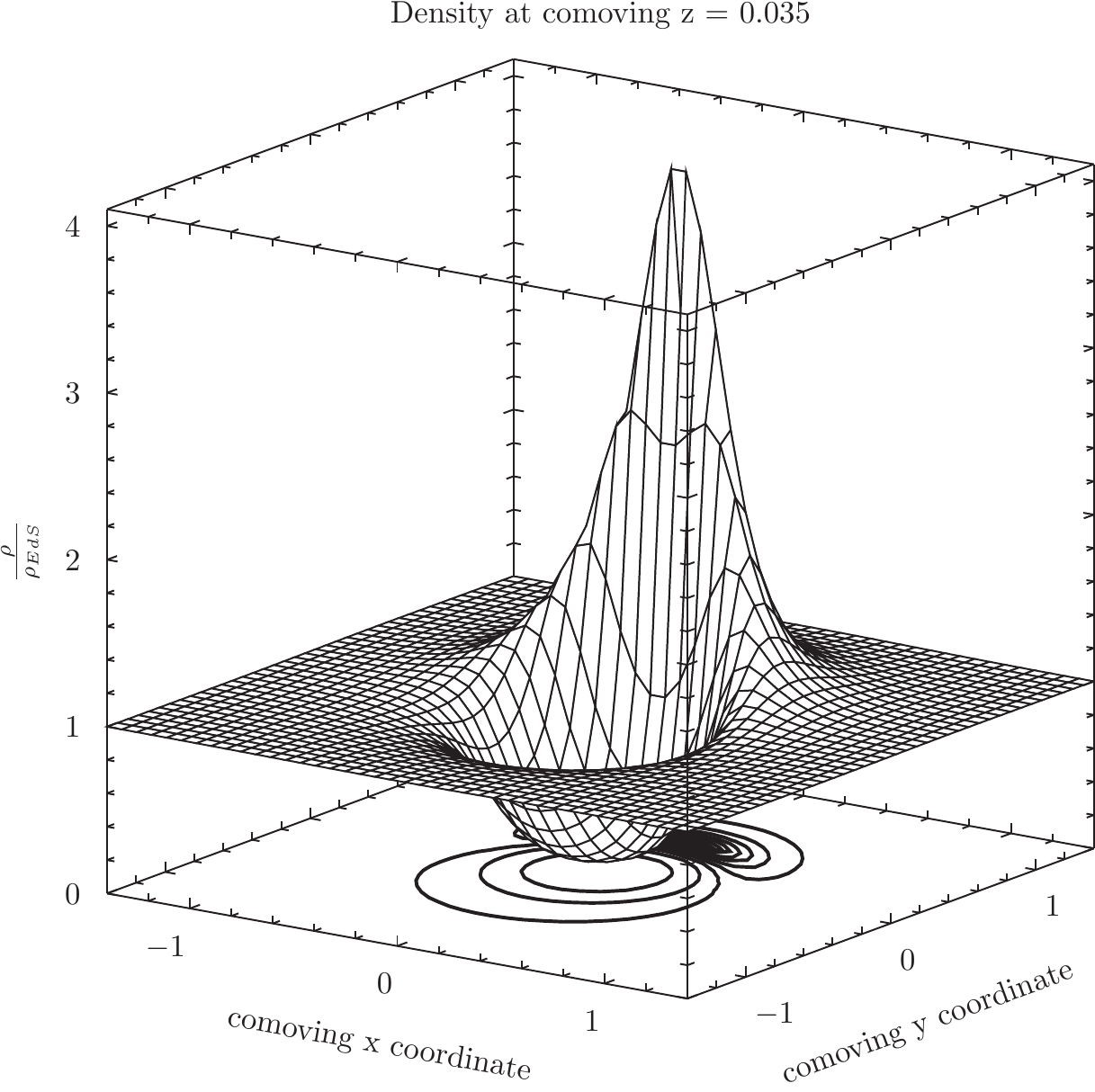}}
\caption{Present time 2D density profile of the Szekeres model. The comoving coordinates are normalized at $t = t_{ls}$ in units of $0.1Mpc$.}
\label{density_2d}
\end{figure}

The procedure that has been used for obtaining specific models follows that introduced by K. Bolejko and described in {\em e.g} \cite{bolejko1}. This method starts by specifying an LTB model with a density distribution that is later altered through appropriate choices of the dipole functions $\tilde P, S$ and $Q$ to create non-symmetric models. Below, a vanishing cosmological constant and $k(r)\leq 0$ is assumed.
\newline \newline
Equation (\ref{eq:friedmann}) can be written as an integral equation:
\begin{equation} \label{bigbangtime}
ct-ct_b(r) = \int_0^A\!\frac{d\tilde A}{\sqrt{2M/\tilde A-k}}\
\end{equation}
Here, only models with a constant time of the big bang will be used, and the constant value will be set equal to zero, {\em i.e.} $t_b(r) = 0$. Introducing a parameter $\eta$, the solution to the integral equation can then be written in parametric form as: 
\begin{equation}\label{A_of_etha}
\begin{split}
A = \frac{M}{-k}(\cosh(\eta)-1),\,\,t = (\sinh(\eta)-\eta)\frac{M}{c(-k)^{3/2}}
\end{split}
\end{equation}
\newline
In the following, the term "background model"  is used for denoting the FLRW model that the LTB model tends to at $r\rightarrow \infty$. In the models studied here, the background model is the Einstein de-Sitter (EdS) model.
\newline\newline
The coordinate covariance in $r$ is eliminated by setting  $A(t_{ls}, r) = r$, where $t_{ls}$ is the time of last scattering in the background model determined by a background redshift $z$ of $z = 1100$.
\newline\indent
To specify the LTB model completely, only one more function is needed. Here, that function is chosen to be $M(r)$. In order to specify $M$, the function is split into two parts by writing $M = \delta M(r)+ M_0$. The latter term corresponds to a background term such that $M_0 = \frac{\Omega_mH_0^2r^3}{2c^2}$, where $\Omega_m$ is the time dependent density parameter of the background model with density $\rho_{EdS}$. $\delta M$ is determined by choosing the desired initial density field. This is done by splitting the density into a background part and a "perturbation" and writing equation (\ref{eq:density}) in the LTB limit as:
\begin{equation}
\rho_{EdS}+\delta\rho = \frac{2((M_0)_{,r}+\delta M_{,r})}{c^2\beta A^2A_{,r}}
\end{equation}
Since $M$ is time-independent, this equation can be integrated at $t = t_{ls}$ to obtain $M$ once $\delta\rho(t_{ls},r)$ has been specified. When this is done, equation (\ref{A_of_etha}) can be solved for $k(r)$ at $t = t_{ls}$ and afterwards it can be solved at any time $t$ to obtain $A(t,r)$.
\newline \newline
The models used here are specified by $\delta\rho(r,t_{ls}) = -10^{-3}0.5\beta\rho_{EdS}(t_{ls})\alpha e^{\frac{-r^2}{\sigma^2(0.1Mpc)^2}}$  with $\alpha = 5.5$, $\sigma = 0.3$ and $\beta$ as defined in equation (\ref{eq:density}). This corresponds to a void with an approximate present time diameter of $160Mpc$. The present time void profile in physical coordinates can be fitted well to the formula given in \cite{void_profile} which describes spherically averaged void profiles according to Newtonian N-body simulations (based on a $\Lambda$CDM background though).
\newline\newline
Aside from the LTB model, also a Szekeres model with no symmetries has been studied. This non-symmetric model is constructed by using the dipole functions defined by $S = 1$, $\tilde P = \text{const.}$ and $Q_{,r} = -8e^{-3r/(0.1Mpc)}\log(r/(0.1Mpc)+1)$. This choice leads to an overdensity peaking with a density contrast $\frac{\rho-\rho_{EdS}}{\rho_{EdS}}$ of approximately $4$ at present time, $t = t_0$.
\newline\indent
2D present time density profiles of the LTB and Szekeres models are shown in figures \ref{ltb_density_2d} and \ref{density_2d}.

\subsection{Geodesic equations and the Angular diameter distance formula in spherical coordinates}\label{section:Szekeres_geodesics}
A complete set of ODEs appropriate for studying redshift-distance relations in Szekeres models was derived in \cite{dallas-light}. The coordinate system used there was the $(t,r,p,q)$ coordinate system. For the purpose of this work it was found convenient to follow the procedure presented in \cite{dallas-light} and re-derive the equations for quasi-spherical Szekeres models in spherical coordinates.
\newline\newline
The line element of the quasi-spherical Szekeres model in spherical coordinates is:
\begin{equation}\label{line_element_sph}
\begin{split}
ds^{2} = -dt^{2}c^{2} + [\frac{(A_{,r}-A\frac{E_{,r}}{E})^2}{(1-k)} +\\
\frac{A^2}{E^2}(S_{,r}^2\cot^2(\frac{\theta}{2}) +2S_{,r}\cot(\frac{\theta}{2})(Q_{,r}\sin(\phi)+\\
\tilde P_{,r}\cos(\phi)) +\tilde P_{,r}^2+Q_{,r}^2)]dr^2 +2\frac{A^2}{E^2}S\cot(\frac{\theta}{2})[Q_{,r}\cos(\phi)-\\
\tilde P_{,r}\sin(\phi)] drd\phi -2\frac{A^2}{E} [Q_{,r}\sin(\phi)+\tilde P_{,r}\cos(\phi)+\\
S_{,r}\cot(\frac{\theta}{2})]drd\theta +A^2d\theta^2 +A^2\sin^2(\theta)d\phi^2
\end{split}
\end{equation}
The line element written in this form can also be found in {\em e.g} \cite{bolejko1}. For convenience, a simplifying notation for the metric functions is used in the following and the ODEs will be given in a notation corresponding to the line element written as:
\newline
\begin{equation}
\begin{split}\label{line_element_short}
ds^{2} = -dt^{2}c^{2}+R(t,r,\theta,\phi)dr^2+2\Phi(t,r,\theta,\phi) drd\phi+\\
2\Theta(t,r,\theta,\phi) drd\theta +F(t,r)d\theta^2+P(t,r,\theta)d\phi^2
\end{split}
\end{equation}
The definition of the metric functions $R,\Phi,\Theta, F, P$ is seen by comparing equations (\ref{line_element_sph}) and (\ref{line_element_short}).
In spherical coordinates $E$ is given by $E = \frac{S}{2\sin^2(\theta/2)}$ while $\frac{E_{,r}}{E} = -\frac{S_{,r}\cos(\theta)+\sin(\theta)[\tilde P_{,r}\cos(\phi)+Q_{,r}\sin(\phi)]}{S}$.
\newline \newline
Using the above metric it is straight forward to obtain the geodesic equations in the spherical coordinate system:
\begin{equation}\label{eq:geod0}
\dot k^t+\frac{1}{2c^2}[R_{,t}(k^r)^2 +F_{,t}(k^{\theta})^2
+P_{,t}(k^{\phi})^2+2\Phi_{,t}k^{\phi}k^r+2\Theta_{,t} k^{\theta}k^r]=0
\end{equation}
\begin{equation}\label{eq:geod1}
\begin{split}
R\dot k^r+\dot R k^r + \dot \Phi k^{\phi} + \Phi \dot k^{\phi} +\dot \Theta k^{\theta} +\Theta \dot k^{\theta}-
\frac{1}{2}[R_{,r}(k^r)^2+\\
F_{,r}(k^{\theta})^2+ P_{,r}(k^{\phi})^2+ 2\Theta_{,r}k^{\theta}k^r+ 2\Phi_{,r}k^{\phi}k^r]=0
\end{split}
\end{equation}
\begin{equation}\label{eq:geod2}
\begin{split}
F\dot k^{\theta}+\dot Fk^{\theta} +\dot \Theta k^{r} +\Theta \dot k^r
- \frac{1}{2}[R_{,\theta}(k^r)^2+\\
P_{,\theta}(k^{\phi})^2+ 2\Theta_{,\theta}k^{\theta}k^r  +2\Phi_{,\theta}k^{\phi}k^r]=0
\end{split}
\end{equation}
\begin{equation}\label{eq:geod3}
\begin{split}
P\dot k^{\phi}+\dot P k^{\phi} + \dot \Phi k^r + \Phi \dot k^r -\frac{1}{2}[R_{,\phi}(k^r)^2+\\ 2\Theta_{,\phi}k^{\theta}k^r+2\Phi_{,\phi}k^{\phi}k^r]=0
\end{split}
\end{equation}
\newline \newline
As with the Newtonian gauge metric, these equations must be differentiated in order to obtain ODEs for $k^t_{,t}, k^r_{,r}, k^{\theta}_{,\theta}$ and $k^{\phi}_{,\phi}$ which are needed in the expression for $D_A$. The resulting equations are:
\begin{equation}\label{eq:time}
\begin{split}
\frac{d}{d\lambda}(k^t_{,\alpha}) + k^t_{,\beta} k^{\beta}_{,\alpha}+\frac{1}{c^2}[\frac{1}{2} R_{,t\alpha} (k^r)^2+\\
R_{,t} k^r k^r_{,\alpha} +
\frac{1}{2} F_{,t\alpha} (k^{\theta})^2+
F_{,t} k^{\theta} k^{\theta}_{,\alpha}+\\
\frac{1}{2} P_{,t\alpha} (k^{\phi})^2+P_{,t}k^{\phi}k^{\phi}_{,\alpha}+
\Phi_{,t\alpha}k^rk^{\phi}+
\Phi_{,t}(k^{\phi}_{,\alpha}k^r+\\
k^{\phi}k^r_{,\alpha})+
\Theta_{,t\alpha}k^{\theta}k^r+\Theta_{,t}(k^{\theta}_{,\alpha}k^r+k^{\theta}k^r_{,\alpha})] =0
\end{split}
\end{equation}
\begin{equation}\label{eq:radius}
\begin{split}
R_{,\alpha}\dot k^r+R(\frac{d}{d\lambda}(k^r_{,\alpha}) +k^r_{,\beta}k^{\beta}_{,\alpha}) +\dot R k^r_{,\alpha}+\\ k^r(R_{,\alpha\beta} k^{\beta}+R_{,\beta}k^{\beta}_{,\alpha})+ k^{\phi}(\Phi_{,\alpha\beta}k^{\beta}+
\Phi_{,\beta}k^{\beta}_{,\alpha}) +\\
\Phi(\frac{d}{d\lambda}(k^{\phi}_{,\alpha} )+k^{\phi}_{,\beta}k^{\beta}_{,\alpha}) + \Phi_{,\alpha} \dot k^{\phi} +\\ 
\dot \Phi k^{\phi}_{,\alpha} +k^{\theta}(\Theta_{,\alpha\beta}k^{\beta}+
\Theta_{,\beta}k^{\beta}_{,\alpha})+
\Theta(\frac{d}{d\lambda}(k^{\theta}_{,\alpha})+\\ k^{\theta}_{,\beta}k^{\beta}_{,\alpha}) + \Theta_{,\alpha} \dot k^{\theta} + \dot \Theta k^{\theta}_{,\alpha} -
[\frac{1}{2} R_{,r\alpha} (k^r)^2+\\
R_{,r}k^rk^r_{,\alpha}+
\frac{1}{2}F_{,r\alpha}(k^{\theta})^2+F_{,r}k^{\theta}k^{\theta}_{,\alpha}+\frac{1}{2} P_{,r\alpha}(k^{\phi})^2 +\\ 
P_{,r} k^{\phi}k^{\phi}_{,\alpha} + \Theta_{,r\alpha}k^{\theta}k^r + \Theta_{,r}(k^{\theta}_{,\alpha}k^r+
k^{\theta} k^r_{,\alpha})+\\
\Phi_{,r\alpha}k^{\phi}k^r+\Phi_{,r} (k^{\phi}_{,\alpha}k^r+k^{\phi}k^r_{,\alpha})] = 0
\end{split}
\end{equation}
\begin{equation}\label{eq:theta}
\begin{split}
F_{,\alpha} \dot k^{\theta} + F(\frac{d}{d\lambda} (k^{\theta}_{,\alpha}) + k^{\theta}_{,\beta} k^{\beta}_{,\alpha}) +
k^{\theta}(F_{,\alpha\beta} k^{\beta} + F_{,\beta} k^{\beta}_{,\alpha}) +\\ \dot F k^{\theta}_{,\alpha} +
k^r(\Theta_{,\beta\alpha} k^{\beta}+ \Theta_{,\beta} k^{\beta}_{,\alpha}) + \dot \Theta k^r_{,\alpha} + \Theta_{,\alpha} \dot k^r +\\
\Theta (\frac{d}{d\lambda}(k^r_{,\alpha}) + k^r_{,\beta}k^{\beta}_{,\alpha}) -  [\frac{1}{2}R_{,\theta\alpha} (k^r)^2 +R_{\theta} k^r k^r_{,\alpha} +\\
\frac{1}{2} P_{,\theta\alpha}(k^{\phi})^2 +
P_{,\theta}k^{\phi} k^{\phi}_{,\alpha} +
\Theta_{,\theta\alpha}k^{\theta}k^r + \Theta_{,\theta} (k^{\theta}_{,\alpha}k^r +\\ 
k^{\theta}k^r_{,\alpha}) + \Phi_{,\theta\alpha} k^{\phi}k^r + \Phi_{,\theta} (k^{\phi}_{,\alpha} k^r+k^{\phi}k^r_{,\alpha})] = 0
\end{split}
\end{equation}
\begin{equation}\label{eq:phi}
\begin{split}
P_{,\alpha} \dot k^{\phi} + P(\frac{d}{d\lambda}(k^{\phi}_{,\alpha}) + k^{\phi}_{,\beta} k^{\beta}_{,\alpha}) +\\ 
k^{\phi} (P_{,\beta\alpha}k^{\beta}+P_{,\beta}k^{\beta}_{,\alpha}) + \dot Pk^{\phi}_{,\alpha} + k^r(\Phi_{,\alpha\beta}k^{\beta} + \Phi_{,\beta}k^{\beta}_{,\alpha}) +\\ 
\dot \Phi k^r_{,\alpha} + \Phi_{,\alpha} \dot k^r + \Phi(\frac{d}{d\lambda}(k^r_{,\alpha}) + k^r_{,\beta} k^{\beta}_{,\alpha}) -\\
[\frac{1}{2} R_{,\phi\alpha} (k^r)^2+
R_{,\phi} k^rk^r_{,\alpha} + \Theta_{,\phi\alpha} k^{\theta}k^r + \Theta_{,\phi} (k^{\theta}_{,\alpha}k^r + k^{\theta}k^r_{,\alpha}) +\\ 
\Phi_{,\alpha\phi} k^{\phi}k^r + \Phi_{,\phi}(k^{\phi}_{,\alpha}k^r + k^{\phi}k^r_{,\alpha})] =0
\end{split}
\end{equation}
Inserting the appropriate Christoffel symbols (given in appendix \ref{Szekeres-appendix}) into equation (\ref{eq:DAalpha}), the following differential equation for the angular diameter distance is obtained:
\begin{equation}\label{eq:DA}
\begin{split}
4\frac{d\ln D_A}{d\lambda} = 2(k^{t}_{,t}+ k^{r}_{,r} +k^{\theta}_{,\theta}+k^{\phi}_{,\phi})+\\ 
\frac{F_{,t}}{F}k^{t}+ \frac{F_{,r}}{F}k^{r} +\frac{P_{,t}}{P}k^{t}+ \frac{P_{,r}}{P}k^{r} +\frac{P_{,\theta}}{P}k^{\theta} +\\
\frac{1}{R-\frac{\Phi^2}{P} -\frac{\Theta^2}{F}}(R_{,t}k^{t} + R_{,r}k^{r} + R_{,\theta}k^{\theta} + R_{,\phi}k^{\phi}) +\\ \frac{1}{F}\frac{\Theta}{R-\frac{\Phi^2}{P} -\frac{\Theta^2}{F}} (k^t(\frac{\Theta}{F}F_{,t}-2\Theta_{,t})+\\
k^r(\frac{\Phi}{P}\Phi_{,\theta}-\frac{\Phi}{P}\Theta_{,\phi}+\frac{\Theta}{F}F_{,r}-2\Theta_{,r})+k^{\theta}(-2\Theta_{,\theta})+\\ k^{\phi}(\frac{\Phi}{P}P_{,\theta}-2\Theta_{,\phi}))+
\frac{1}{P}\frac{\Phi}{R-\frac{\Phi^2}{P} -\frac{\Theta^2}{F}} (k^t(\frac{\Phi}{P}P_{,t}-\\
2\Phi_{,t})+k^r(\frac{\Theta}{F}\Theta_{,\phi}-\frac{\Theta}{F}\Phi_{,\theta}+\frac{\Phi}{P}P_{,r}-2\Phi_{,r})\\
+k^{\theta}(\frac{\Phi}{P}P_{,\theta}-2\Phi_{,\theta})+k^{\phi}(-2\Phi_{,\phi}-\frac{\Theta}{F}P_{,\theta}))
\end{split}
\end{equation}
This equation is solved simultaneously with the 24 ODEs for the $k^{\alpha}$'s and their derivatives.
\newline \newline 
The last set of equations needed are again the null-condition and its partial derivatives. These equations are used when setting the initial conditions and for checking the accuracy of the code.
\newline \indent
In spherical coordinates, the null-condition is given by:
\begin{equation}
\begin{split}
k^{\alpha}k_{\alpha} = k^{\alpha}k^{\beta}g_{\alpha\beta} = -c^2(k^t)^2+R(k^r)^2+\\
F(k^{\theta})^2+P(k^{\phi})^2+2\Theta k^rk^{\theta}+ 2\Phi k^r k^{\phi} = 0
\end{split}
\end{equation}
Taking the partial derivative of this equation one obtains:
\begin{equation}\label{eq:partialnull}
\begin{split}
-2c^2k^tk^t_{,\alpha} + R_{,\alpha}(k^r)^2 + 2Rk^rk^r_{,\alpha} + F_{,\alpha}(k^{\theta})^2 +\\ 
2F k^{\theta}k^{\theta}_{,\alpha} + P_{,\alpha}(k^{\phi})^2 +2Pk^{\phi}k^{\phi}_{,\alpha}
+2 \Theta_{,\alpha}k^r k^{\theta} +\\ 
2\Theta(k^r_{,\alpha} k^{\theta} + k^r k^{\theta}_{,\alpha}) +
2\Phi_{,\alpha}k^r k^{\phi} + 2\Phi (k^r_{,\alpha}k^{\phi} + k^rk^{\phi}_{,\alpha}) = 0
\end{split}
\end{equation}
\newline\newline
The equations given above are shown in expanded form in appendix \ref{Szekeres-appendix} together with comments on initial conditions.

\subsection{Relation between the Szekeres and Newtonian gauge spacetimes}\label{section_mapping}
In this section, the subscripts "$sz$" denotes Szekeres coordinates including LTB coordinates while the subscript $"ltb"$ is used for specifying LTB coordinates. The subscript "$ng$" denotes Newtonian gauge coordinates and the subscript "$flrw$" is used for denoting coordinates on the unperturbed FLRW background. A tilde will be used to denote coordinates of fiducial spacetime points.
\newline\newline
The numerical values of coordinates are of no physical value. Instead, the physically relevant quantities are proper distances, and these are determined by the metric. In order for two spacetimes to be considered equivalent, a map between their coordinates must thus be constructed such that proper distances are equal in the two spacetimes. It could seem, that the appropriate point identification map in this case is between the Szekeres and Newtonian gauge spacetimes and this would indeed also lead to an interesting study. However, the recipe of \cite{recipe} is for going between the Newtonian gauge metric and an N-body simulation with its underlying FLRW metric. The intent here is to see how well the recipe describes the "true" underlying relativistic spacetimes corresponding to density and velocity fields obtained from N-body simulations. This is done by studying the Newtonian gauge metric's ability to accurately describe light propagation. Such a study requires the knowledge of the "true" underlying spacetime of the considered N-body fields. Thus, mock N-body data is constructed by mapping Szekeres models into their background FLRW models. The mapped Szekeres model's density profile and the velocity field obtained through the mapping comprise mock N-body data. As in regular perturbation theory, the Newtonian gauge metric is then assigned the same coordinate system as that of the FLRW background of the "N-body data".
\newline\indent
Both the velocity and density fields of LTB models can be reproduced by N-body simulations when using initial conditions based on the maps described below. As mentioned in the introductory section, it is still work in progress to show that this is also the case for non-symmetric Szekeres models. Following the standard consensus, it is here assumed that Newtonian N-body simulations accurately reproduce structure formation in accordance with general relativity. In particular, it is thus assumed that non-symmetric quasi-spherical Szekeres models {\em are} reproducible by Newtonian N-body simulations.
\newline\newline
Following \cite{Troels}, the point identification map between FLRW and LTB coordinates is defined by the requirement of equal proper radial distances in the two spacetimes. Letting $g_{\alpha\beta}$ denote components of the metric tensor, the identification between comoving coordinates in the two spacetimes is thus given by the four equations:
\begin{equation}\label{map}
\begin{split}
\tilde t:=\tilde t_{ltb} = \tilde t_{flrw}\\
\tilde\theta:= \tilde\theta_{ltb} = \tilde\theta_{flrw}\\
\tilde\phi:=\tilde\phi_{ltb} = \tilde\phi_{flrw}\\
dp_{r,ltb} :=\int_{0}^{\tilde r_{ltb}}dr_{ltb}\sqrt{g_{rr,ltb}} = \int_{0}^{\tilde r_{flrw}}dr_{flrw}\sqrt{g_{rr,flrw}}
\end{split}
\end{equation}
These equations yield a mapping $(\tilde t, \tilde r_{flrw}, \tilde \theta, \tilde\phi  ) \rightarrow(\tilde t, \tilde r_{ltb}, \tilde \theta, \tilde\phi  )$. Note that since the big bang time in equation (\ref{bigbangtime}) is zero, the time coordinates of the two spacetimes are the same. Note also, that the spherical symmetry about the origin of the LTB model implies that the angular coordinates of the LTB and FLRW metric are the same.
\newline\newline
The non-symmetric Szekeres metric is non-diagonal in spherical coordinates while the FLRW metric is diagonal. A point identification map in spherical coordinates does thus not capture the complete anisotropy of the Szekeres model. As shown in equation (\ref{eq:line_elemet}) the general Szekeres metric is diagonal in stereographic coordinates defined by equation (\ref{stereo}). The appropriate point identification map between the non-symmetric Szekeres spacetime and the FLRW spacetime is thus defined in stereographic coordinates:
\begin{equation}\label{stereo_map}
\begin{split}
\tilde t:=\tilde t_{sz} = \tilde t_{flrw}\\
\tilde p:= \tilde p_{sz} = \tilde p_{flrw}\\
\tilde q:=\tilde q_{sz} = \tilde q_{flrw}\\
dp_{r,sz} := \int_{0}^{\tilde r_{sz}}dr_{sz}\sqrt{g_{rr,sz}} = \int_{0}^{\tilde r_{flrw}}dr_{flrw}\sqrt{g_{rr,flrw}}
\end{split}
\end{equation}
In this equation, $g_{rr,sz}$ is the $rr$-component of the Szekeres metric in stereographic coordinates while $g_{rr,ltb}$ was the $rr$-component of the LTB metric in spherical coordinates. The $rr$-component of the FLRW metric is the same in spherical and stereographic coordinates as the flat FLRW metric in stereographic coordinates is given by:
\begin{equation}
ds^2 = -c^2dt_{flrw}^2 + a^2\left( dr_{flrw}^2 + \frac{r_{flrw}^2}{\tilde E^2}(dp_{flrw}^2 + dq_{flrw}^2)\right) 
\end{equation}
The function $\tilde E$ in this line element is defined by $\tilde E = \frac{1}{2}\left( p_{flrw}^2 + q_{flrw}^2 +1 \right) $ which in spherical coordinates corresponds to $\tilde E = \frac{1}{2\sin^2(\theta_{flrw}/2)}$. The stereographic FLRW coordinates are related to the spherical FLRW coordinates by the transformation:
\begin{equation}\label{flrw_stereo}
\begin{split}
p_{flrw} = \cot\left(\theta_{flrw}/2\right) \cos(\phi_{flrw})\\
q_{flrw} = \cot\left(\theta_{flrw}/2\right) \sin(\phi_{flrw})
\end{split}
\end{equation}
\newline\newline
Combining the stereographic point identification map with the coordinate transformations of equations (\ref{stereo}) and (\ref{flrw_stereo}), a point identification map between the spherical coordinate systems of the Szekeres and FLRW spacetimes is obtained. The stereographic coordinates of the Szekeres model are related to the angular coordinates through an $r$-dependence but this is not the case for the FLRW metric. This difference implies that the angular coordinates in the two spacetimes will not be identical.
\newline\indent
Other point identification maps than the ones given above have been studied. In particular, a map requiring equal proper distances along the $(p,q)$-coordinate directions instead of $p_{sz}=p_{flrw}, q_{sz}=q_{flrw}$ has been studied along with two maps in spherical coordinates. For the non-symmetric Szekeres model, the maps do not yield the same results indicating a lack of exact compatibility between the FLRW and general Szekeres spacetimes. The map shown in equation (\ref{stereo_map}) is used because it yields the best reproduction of the non-symmetric Szekeres spacetime.
\newline\newline

\paragraph{Peculiar velocities and $\psi$:}
In the Szekeres model, the observer is comoving and the redshift is given by $1+z = \frac{k^t(\lambda)}{k^t_0}$, with a subscript $0$ referring to evaluation at the position of the present time observer. In the Newtonian gauge spacetime, dust is generally not comoving. Following \cite{recipe}, the appropriate peculiar velocity field is proportional to $(1,v^i)$ with $v^i = v_i$ the local comoving spatial dust motion on the EdS background/in the N-body simulation. The velocity field is normalized such that $u^{\mu}u_{\mu} = -c^2$ so the appropriate field is $u^{\mu} = (1,v^i)\frac{c}{\sqrt{c^2(1+2\psi) - a^2(1-2\psi)(v_r^2+ v_{\theta}^2r^2 + v_{\phi}^2r^2\sin^2(\theta))}}$. The velocity field is needed in order to obtain the redshift along the geodesics using the general formula $1+z = \frac{(k^{\alpha}u_{\alpha})_e}{(k^{\alpha}u_{\alpha})_0}$, with a subscript $e$ indicating the spacetime position of emission {\em i.e.} $k^{\alpha}_e = k^{\alpha}(\lambda)$ etc.
\newline\indent
The comoving peculiar radial velocity $v_r$ is computed by taking the time derivative of the expression for the proper distance $dp_r$:
\begin{equation}
\begin{split}
\frac{d}{dt}dp_r = \frac{d}{dt}(ar_{flrw}) = a_{,t}r_{flrw} + av_{r}\\
\end{split}
\end{equation}
The proper distance on the left hand side is computed in the Szekeres/LTB spacetime at the appropriately mapped spacetime point.
\newline\indent
The angular velocities vanish in the LTB model because of the spherical symmetry about the origin. The angular velocities of the non-symmetric Szekeres model do not vanish identically, but they are small and are only used in the formula for the redshift in which they are suppressed by $k^{\theta}_{ng}, k^{\phi}_{ng}$. The angular velocities can thus be neglected.
\newline\newline
The potential $\psi_{ng}$ is needed in order to solve the geodesic equations and for obtaining $D_A$ in the Newtonian gauge. The potential is obtained by the usual Poisson equation $\nabla^2\psi_{ng} = \frac{4\pi G a^2}{c^2}\delta\rho_{ng}$. This equation is solved in Fourier space (using fftw3\footnote{http://www.fftw.org/}) on a grid and the value of $\psi_{ng}$ at a specific point is then obtained using quadri-linear interpolation. This method should approximately mimic how one would work with the potential from an N-body simulation.
\newline\indent
The overdensity is defined as $\delta\rho := \rho_{Szekeres}-\rho_{EdS}$ and is obtained at spacetime points in the Szekeres metric. The overdensity needed for obtaining $\psi_{ng}$ is the corresponding overdensity of the mock N-body simulation so a mapping into the Newtonian gauge spacetime is necessary. The overdensity $\delta\rho_{ng}$  at a Newtonian gauge spacetime point $(\tilde t_{ng}, \tilde r_{ng}, \tilde \theta_{ng},\tilde \phi_{ng}) = (\tilde t_{flrw}, \tilde r_{flrw}, \tilde \theta_{flrw}, \tilde\phi_{flrw})$ is thus computed as $\delta\rho_{ng}(\tilde t_{ng}, \tilde r_{ng}, \tilde \theta_{ng}, \tilde \phi_{ng}) = \delta\rho_{Szekeres}(\tilde t_{sz}, \tilde r_{sz}, \tilde \theta_{sz}, \tilde \phi_{sz})$, where $(\tilde t_{sz}, \tilde r_{sz}, \tilde \theta_{sz}, \tilde \phi_{sz})$ is the Szekeres spacetime point corresponding to $(\tilde t_{ng}, \tilde r_{ng}, \tilde \theta_{ng}, \tilde \phi_{ng})$ according to the point identification map defined by equation (\ref{map}) or (\ref{stereo_map}). The corresponding potential in the Newtonian gauge spacetime, $\psi_{ng}$, is then computed from the Poisson equation in the Newtonian gauge spacetime.

\subsection{Gauge transformation of the angular diameter distance}
In the previous subsection, it was discussed how to construct a point identification map between the Szekeres model and its FLRW background in order to construct mock N-body data. This map is used for obtaining the metric potential $\psi$ and peculiar velocities of the Newtonian gauge metric so that this metric describes the Szekeres spacetime. Another issue regarding the use of the two spacetimes is that they do not correspond to the same spacetime slicing since the Szekeres metric is given in the comoving synchronous slicing of spacetime. In order to compare $D_A(z)$ obtained from the Newtonian gauge equations with the exact result, it must thus in principle undergo a gauge transformation to the comoving synchronous gauge.
\newline\indent
It is well known from linear relativistic perturbation theory that the differences in observable quantities in different gauges are negligible well within the horizon. The structures studied here have a present time diameter of approximately $160Mpc$ so the gauge transformation should not be necessary. A formal gauge transformation is performed anyway as a precaution.
\newline\newline
Redshift perturbations are gauge dependent, but the redshift itself is a scalar {\em i.e.} coordinate invariant. Since the value of the exact redshift is computed here, and not just the perturbation, no transformation of the redshift is thus necessary. The angular diameter distance is not gauge invariant; it is the fraction of two infinitesimal areas and areas change under coordinate transformations.
\newline\indent
As will become apparent below, only the transformation of the time coordinate is needed to obtain a gauge transformation of the angular diameter distance. The transformation law between the time coordinates in the two coordinate systems is found by using the general coordinate transformation equation $g_{\mu'\nu'} = g_{\alpha\beta}\frac{ \partial x^{\alpha} }{\partial x^{\mu'}    } \frac{\partial x^{\beta}}{\partial x^{\nu'} } $. Inserting $-c^2$ as the time-time component of the Synchronous gauge metric tensor into the left hand side of this equation and inserting the Newtonian gauge metric into the right hand side, yields the approximate transformation law:
\begin{equation}\label{eq:gauge}
\begin{split} 
t_S   \approx t_N +  \int_0^{t_N}\! \psi dt \
\end{split}
\end{equation}
In this equation, $t_N$ and $t_S$ are the time coordinates in the Newtonian and Synchronous gauges respectively.
\newline\indent
The approximate result is the same as what one would obtain by using regular gauge transformation laws for going between the Newtonian gauge and a synchronous gauge (see {\em e.g.} section 5.3 of \cite{Weinberg}). By using the first order Euler equation in the spherically symmetric limit it is seen that  $-c^2  \int_0^{t_N}\! \psi dt \ = -a\int_r^{\infty}\! v_r(t_N,r') dr' \ $ which shows that the approximate result obtained here is also in agreement with that found in \cite{LTB_in_NG}.
\newline\newline
Using the simplified notation $x^{\alpha}_S = x^{\alpha}_N + \xi^{\alpha}$ for the gauge transformation of $x^{\alpha} = (t,r,\theta,\phi)$, a Taylor expansion reveals the gauge transformation of the angular diameter distance:
\begin{equation}
\begin{split}
D_{A,S}(x^{\alpha}_{e,S}, x^{\alpha}_{0,S}) \approx  D_{A,N}(x^{\alpha}_{e,N}, x^{\alpha}_{0,N}) +\\ 
(\bar D_A(x^{\alpha}_e, x^{\alpha}_0))_{,\alpha}\xi^{\alpha}|_e +(\bar D_A(x^{\alpha}_e,x^{\alpha}_0))_{,\alpha}\xi^{\alpha}|_0
\end{split}
\end{equation}
An over-bar is used for denoting background quantities and coordinates without a subscripted $N$ or $S$ are background coordinates. As before, $N$ and $S$ denote coordinates in the Newtonian and comoving synchronous gauge respectively, and $e$ denotes point of emission while $0$ denotes point of observation.
\newline\newline
Using the Mattig relation \cite{relativistic_cosmology}, $\bar D_A = 2cH_0^{-1}\left[ \frac{a(t_e)}{a(t_0)} -\frac{a(t_e)^{3/2}}{a(t_0)^{3/2}} \right] $, the gauge transformation law becomes:
\begin{equation}
\begin{split}
D_{A,S}(x^{\alpha}_{e,S}, x^{\alpha}_{0,S}) \approx   D_{A,N}(x^{\alpha}_{e,N}, x^{\alpha}_{0,N}) +\\
2cH_0^{-1} \left[ \frac{a_{,t}(t_e)}{a(t_0)} -
\frac{3}{2}\frac{\sqrt{a(t_e)}a_{,t}(t_e)}{a(t_0)^{3/2}}   \right]  \xi^t|_e - c\xi^t|_0
\end{split} 
\end{equation}

\section{Results}\label{results}
The system of ODEs presented in section \ref{section:DA_newtonian} are used to obtain $D_A(z)$ along single geodesics in the LTB and non-symmetric Szekeres models described by the Newtonian gauge metric. Afterwards, the exact $D_A(z)$ relation is obtained by solving the set of ODEs given in section \ref{section:Szekeres_geodesics}. In order to compare the geodesics and the corresponding distance-redshift relations obtained with the two different sets of ODEs, the geodesics must of course be initialized equivalently. Since the equations are solved backwards in time, this implies that the observer position and line of sight must be mapped from the Newtonian gauge coordinate system to Szekeres coordinates. The position of the observer is mapped using the maps of equation (\ref{map}) or (\ref{stereo_map}). The line of sight of the observer is mapped by mapping the Newtonian gauge geodesic into Szekeres spacetime and computing a two point finite difference along the beginning of this geodesic. The two-point finite difference constitutes the initial conditions of $(k^r,k^{\theta}, k^{\phi})$ and the initial condition of $k^t$ is obtained through the null-condition. A geodesic which is initialized as radial at the position of a central observer is radial in both spacetimes. No mapping of $k^r,k^{\theta}, k^{\phi}$ is necessary in this case since the results are frequency independent and $k^t$ and $k^r$ uniquely determine each other through the null-condition. The initial direction of the ray is determined by a set of angular coordinates $(\theta,\phi)$ which must be mapped though.

\subsection{Geodesics in the LTB spacetime}
\begin{figure}
\centerline{\includegraphics[scale = 0.7]{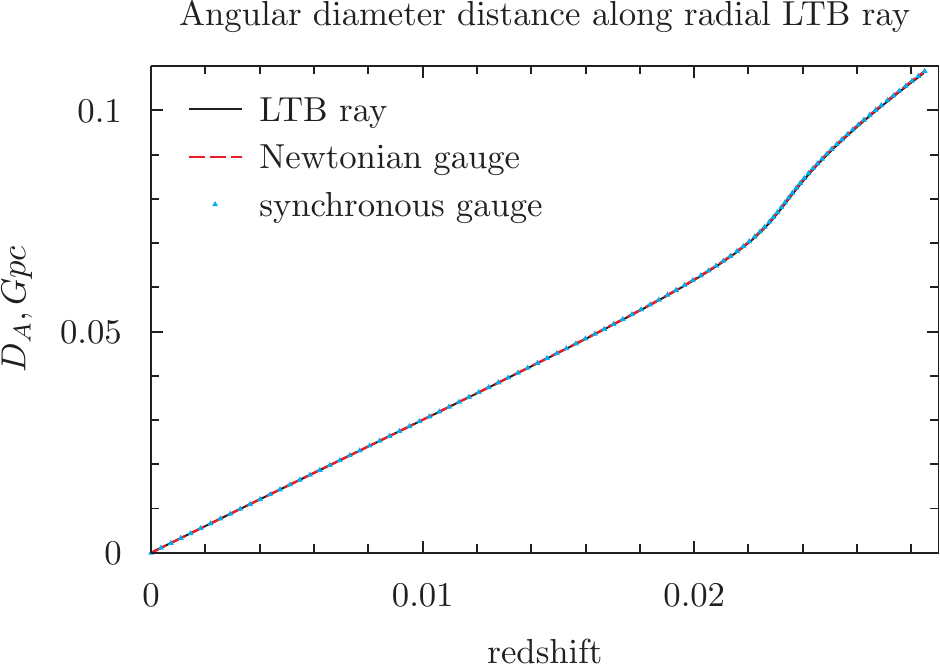}}
\caption{Angular diameter distance along a radial geodesic in the LTB model with a central observer. The line with the legend "LTB ray" is the angular diameter distance along a geodesic computed with the ODE's of the exact LTB metric. The two other lines are the angular diameter distances along the "equivalent" geodesic according to the Newtonian gauge metric in the Newtonian gauge and in the synchronous gauge respectively - these two lines are indistinguishable as expected. They are also indistinguishable from the first mentioned line indicating that the Newtonian gauge metric adequately describes the LTB geodesic.}
\label{ltb_DA}
\end{figure}

\begin{figure}
\centerline{\includegraphics[scale = 0.7]{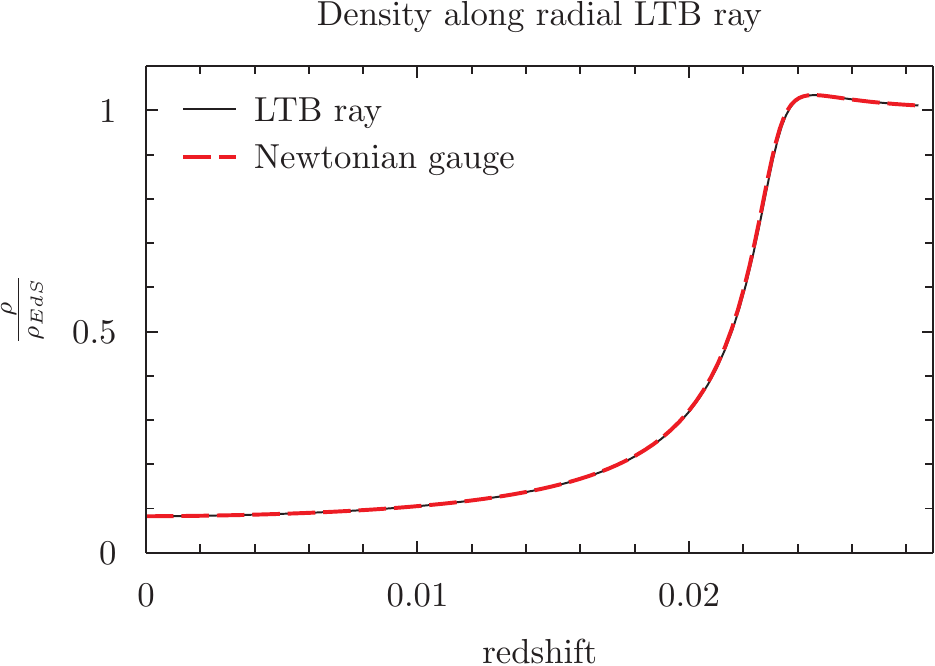}}
\caption{Density along a radial geodesic in the LTB model with a central observer. The density is plotted both according to the Newtonian gauge geodesic and the Szekeres geodesic, but the two densities are the same implying that the two geodesics pass through equivalent portions of spacetime.}
\label{rho_ltb}
\end{figure}

\begin{figure}
\centerline{\includegraphics[scale = 0.7]{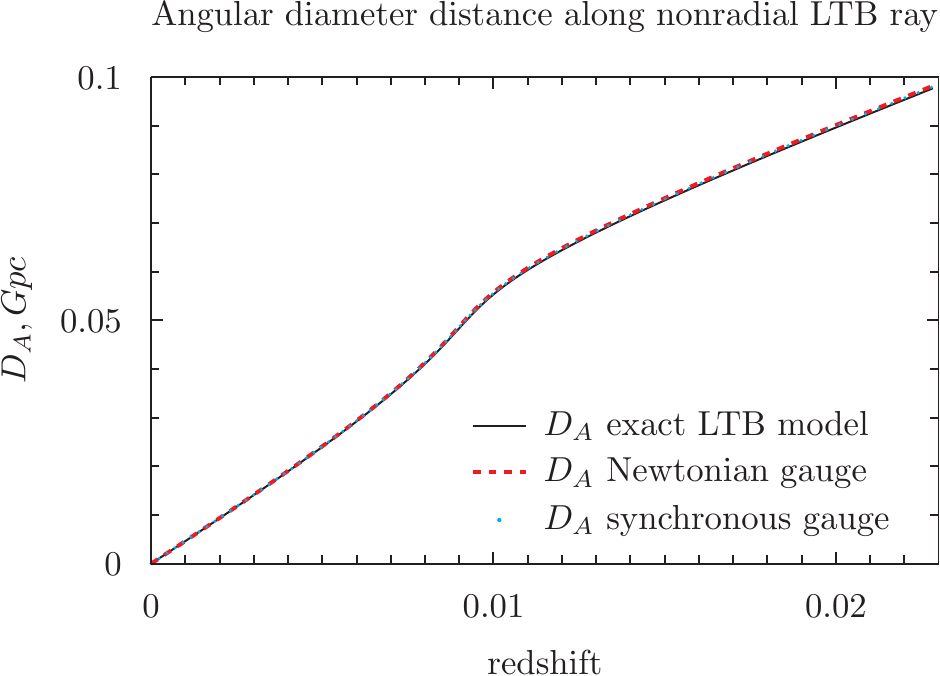}}
\caption{Angular diameter distance along a non-radial geodesic in the LTB model with an observer placed outside the void at $r = 1.2$ in Newtonian gauge coordinates (comoving coordinates are normalized at $t = t_{ls}$ in the units of $0.1Mpc$). The geodesic is initialized in the spatial direction determined by $(k^r_0,k^{\theta}_0, k^{\phi}_0) = (-0.01,0.001,0)$ in the LTB spacetime. The line with the legend "LTB ray" is the angular diameter distance along a geodesic computed with the ODE's of the exact LTB metric. The two other lines are the angular diameter distances along the "equivalent" geodesic according to the Newtonian gauge metric in the Newtonian gauge and in the synchronous gauge - as expected, these two lines are indistinguishable. They are also indistinguishable from the first mentioned line indicating that the Newtonian gauge metric adequately describes this non-radial LTB geodesic.}
\label{ltb_DA_nonrad}
\end{figure}

\begin{figure}
\centerline{\includegraphics[scale = 0.7]{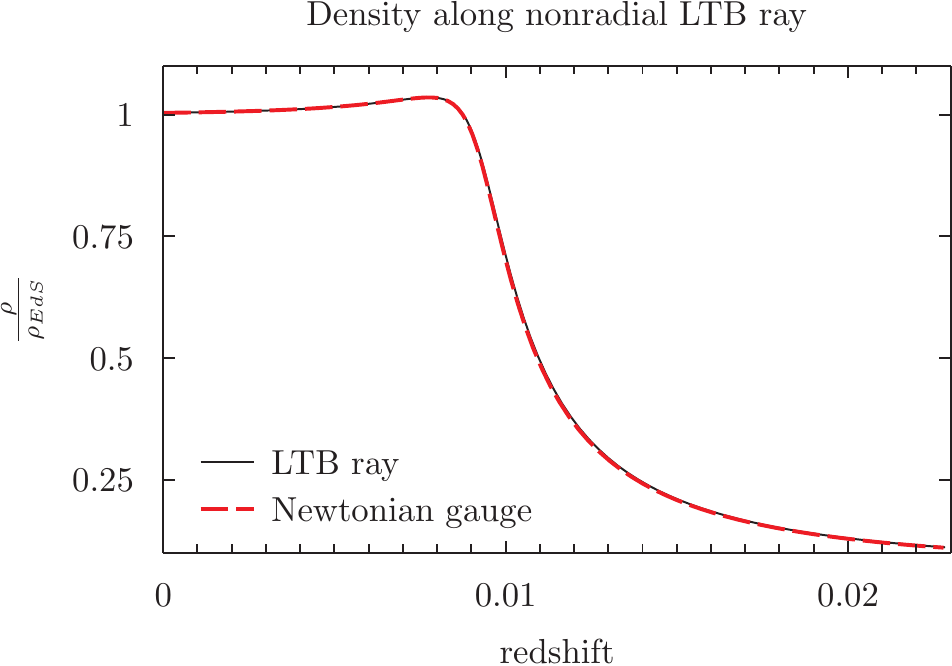}}
\caption{Density along a non-radial geodesic in the LTB model with the observer placed at comoving $r = 1.2$ in Newtonian gauge coordinates (comoving coordinates are normalized at $t = t_{ls}$ in the units of $0.1Mpc$). The observer is looking in the direction determined by $(k^r_0,k^{\theta}_0, k^{\phi}_0) = (-0.01,0.001,0)$ in the LTB spacetime. The density is shown along both the Szekeres and Newtonian gauge geodesics. The two densities are the same though, indicating that the Newtonian gauge adequately reproduces the LTB geodesic.}
\label{rho_nonrad_ltb}
\end{figure}

In figures \ref{ltb_DA} and \ref{ltb_DA_nonrad}, $D_A(z)$ is shown for a radial and a non-radial geodesic in the LTB model respectively. The radial geodesic corresponds to a central observer, while the observer is placed outside the void in the non-radial case. For the radial ray, there are only negligible differences between the curves obtained from solving the ODE system based on the Newtonian gauge metric and that based on the exact LTB metric. For the non-radial ray, there is a slightly more noticeable difference between the distance-redshift relation obtained with the exact metric and that obtained with the Newtonian gauge metric. This difference is presumably due to precision errors occurring when computing the initial values of $k^r,k^{\theta},k^{\phi}$ in one spacetime from their values in the other spacetime.
\newline\indent
In the figures, $D_A(z)$ is also shown after a transformation to the comoving synchronous gauge. Clearly, this transformation is completely obsolete as was expected.
\newline\newline
In figures \ref{rho_ltb} and \ref{rho_nonrad_ltb} the density profiles along the rays are shown
\footnote{Note that in the abridged dictionary of \cite{recipe}, the density used for computing the potential is not the same as the density of the Newtonian gauge spacetime. Using the notation of \cite{recipe}, $\delta = \delta_N-\frac{3}{4\pi G\rho_{EdS}a^2}(a_{,t}^2\psi_N + a_{,t}\psi_{,t})$, where $\delta$ is the density contrast corresponding to the Newtonian gauge spacetime while $\delta_N$ is the density contrast of the N-body simulation and $\psi_N$ the corresponding potential. $\psi_N$ is the potential perturbation appearing in the metric, while $\delta$ describes the density field of the spacetime. In the notation used here, $\delta_N$ corresponds to $\delta_{ng}$ which was used to compute the potential. The difference between the two density fields of the dictionary is insignificant for the models studied here.}
. These are the same along the exact LTB rays and the Newtonian gauge rays which shows that the rays in the two spacetimes follow equivalent spacetime paths.
\newline\indent
It is not surprising that the distance-redshift relation of the LTB model is reproduced by the perturbed FLRW model in the Newtonian gauge, as it has earlier been shown that the Newtonian gauge describes the LTB spacetimes well \cite{LTB_in_NG, LTB_in_NG2, LTB_in_NG3}.

\subsection{Geodesics in non-symmetric Szekeres spacetimes}
\begin{figure}
\centerline{\includegraphics[scale = 0.7]{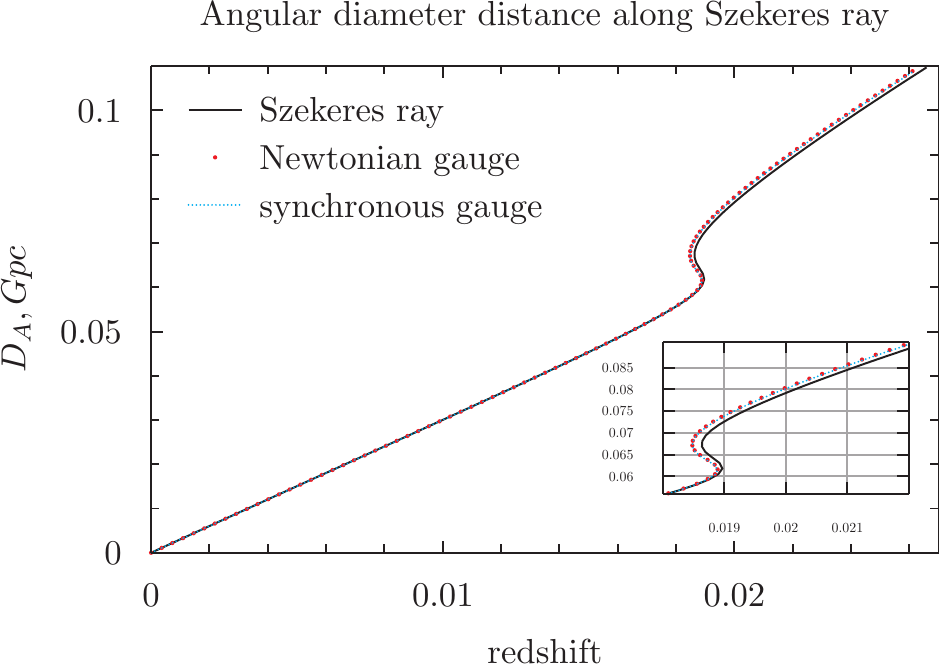}}
\caption{Angular diameter distance along a geodesic in the non-symmetric Szekeres model with a central observer.  The ray is initialized as radial in the direction specified by $(\theta_0, \phi_0) = (\pi/3,\pi/3)$ in Newtonian gauge spacetime coordinates. The line with the legend "Szekeres ray" is the angular diameter distance along a geodesic computed with the ODE's of the exact Szekeres metric. The two other lines are the angular diameter distances along the "equivalent" geodesic according to the Newtonian gauge metric in the Newtonian gauge and in the synchronous gauge. A close-up of the area where the exact and Newtonian gauge results begin to be distinguishable is included.}
\label{sz_DA}
\end{figure}

\begin{figure}
\centerline{\includegraphics[scale = 0.7]{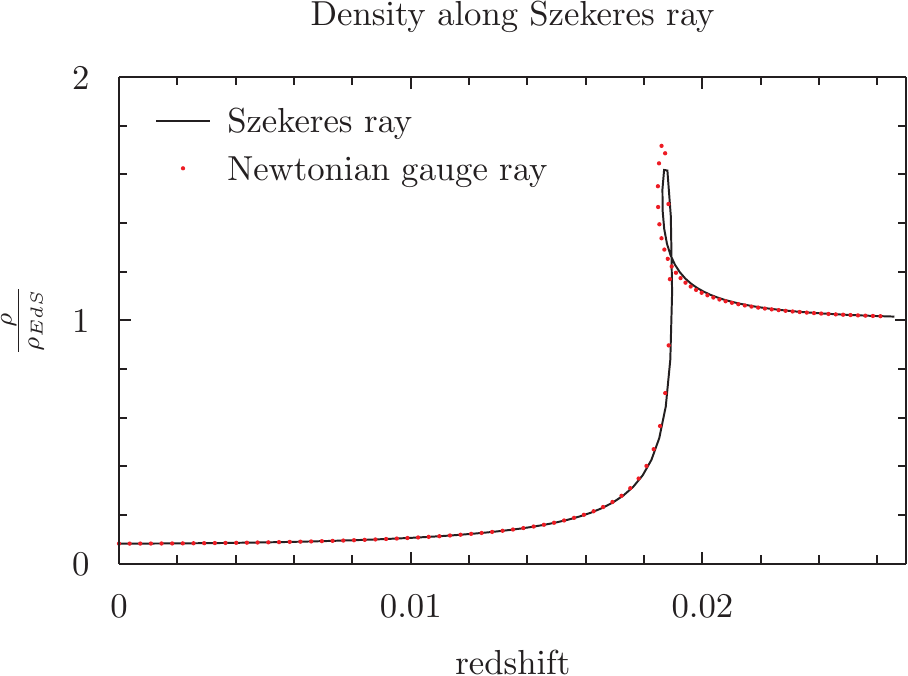}}
\caption{Density along a geodesic in the non-symmetric Szekeres model with a central observer. The ray is initialized at a central observer as a radial ray in the direction specified by $(\theta_0, \phi_0) = (\pi/3,\pi/3)$ in Newtonian gauge spacetime coordinates.}
\label{rho_sz}
\end{figure}

\begin{figure}
\centerline{\includegraphics[scale = 0.7]{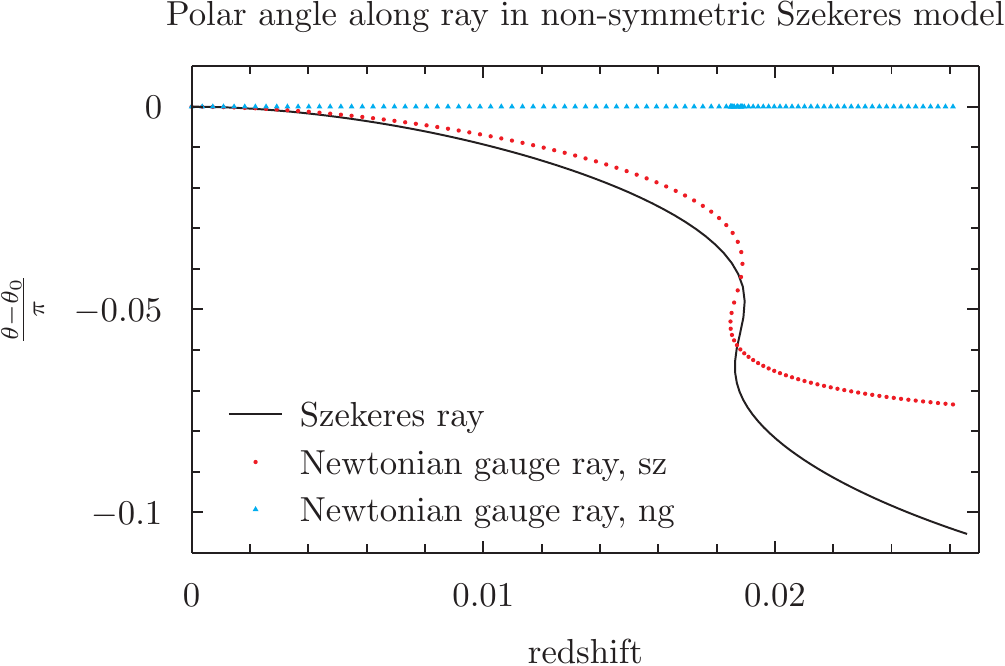}}
\caption{Polar angle along the geodesic in non-symmetric Szekeres model. The angular changes are shown along both the exact ray according to the Szekeres metric ("Szekeres ray") and along the ray according to the Newtonian gauge metric ("Newtonian gauge ray"). For the Newtonian gauge geodesic, the angular changes are shown in both Szekeres and Newtonian gauge coordinates, denoted by "sz" and "ng" respectively.}
\label{sz_theta}
\end{figure}

In figure \ref{sz_DA}, the distance-redshift relation along a geodesic in the non-symmetric Szekeres model is shown. The angular diameter distance is not as precisely reproduced by the Newtonian gauge metric as in the LTB case, but the reproduction is still very good. The difference between the $D_A(z)$ relations along the two rays is consistent with the Newtonian gauge ray moving through slightly larger overdensity than the exact ray. This is in fact the case, as can be seen in figure \ref{rho_sz} which shows the overdensities along the two geodesics. In figure \ref{sz_theta} the polar angle along the two geodesics is shown. The geodesic is not bent in the Newtonian gauge metric, but when mapping the angular coordinates into the Szekeres coordinates, the geodesic is seen to correspond to a non-radial Szekeres ray. The ray is not bent exactly as the exact ray though.
\newline\indent
The slight under-bending of the ray inside the void is consistent with its moving through a larger overdensity at the void edge and implies that the Newtonian gauge ray and the exact ray do not move along exactly equivalent spacetime paths. Similar results are obtained when studying rays in Szekeres models with less anisotropy and smaller density contrasts, with the disagreement between the Newtonian gauge and exact Szekeres results becoming less as the level of anisotropy and overdensity are decreased. Models with more anisotropy have not been studied.
\newline\newline
As with the LTb rays, the gauge dependence of the distance-redshift relation along geodesics in the non-symmetric Szekeres models is completely insignificant.

\section{Conclusions}
The Newtonian gauge metric corresponding to the recipe of \cite{recipe} was used to derive a set of ODEs that can be solved to obtain distance-redshift relations in the corresponding spacetime. The equations were used to obtain the angular diameter distance as a function of the redshift in mock N-body simulations based on quasi-spherical Szekeres models. The validity of the recipe was then estimated by comparing with redshift-distance relations obtained by using the exact Szekeres metric. In the spherically symmetric case, the distance-redshift relation obtained with the Newtonian gauge metric and the exact metric of the model agreed. This is not surprising since others have earlier shown that LTB models after a gauge transformation at least approximately can be described as spherical perturbations on an FLRW background.
\newline\indent
The results obtained here emphasize that it is futile to try to refute results obtained with LTB models by claiming that they are unreliable because of spurious gauge modes in the synchronous gauge; it is a well known result from linear relativistic perturbation theory that observables have negligible gauge dependence on scales below the horizon. Here it has been shown explicitly that this is specifically the case for distance-redshift relations in quasi-spherical Szekeres models - even when the density contrast moves into the non-linear regime.
\newline\newline
For the non-radial geodesic in the non-symmetric Szekeres model, there are small but definite differences between the geodesic paths and distance-redshift relations obtained with the two metrics. The ray appears not to being bent enough inside the void which leads it to move through a slightly larger density contrast than the exact ray. This again induces a slight difference in the distance-redshift relations obtained along the exact and Newtonian gauge geodesics.
\newline\newline
Versions of the studied recipe are the typical basis for ray tracing schemes. Standard ray tracing methods do not include treating the Newtonian gauge metric as exact though, and typically involve several different approximations. For example, a ray is typically not traced by the actual Newtonian gauge geodesics, but is traced in the background FLRW metric, possibly being bent at a finite number of lens planes. This may lead to results that are in less agreement with the exact results than what was obtained here. In addition, the small discrepancies between the exact and Newtonian gauge rays found here may be cumulative though it seems just as likely that the effects will cancel each other if the rays move through several structures instead of just the one they moved through in this work.
\newline\indent
To study how accurate actual ray tracing techniques are in reproducing shear, magnification etc. in exact spacetimes, a study of quasi-spherical swiss cheese Szekeres models is under way. If any shortcomings appear and especially if they are statistically robust, more accurate ray tracing techniques may need to be developed in order to meet the increasingly high precision of observations. In such a case, exact inhomogeneous and anisotropic cosmologically relevant models seem like a good tool for general developments and tests of more accurate ray tracing methods {\em e.g} by studying the use of more advanced metric approximations such as one based on a post-Friedmann expansion (see {\em e.g.} \cite{rayTrace_igen}) or the "Oxford" dictionary in \cite{recipe}.

\section{Acknowledgments}
We thank the anonymous referee for his/her suggestions which have led to significant improvements in the presentation of our work compared to the original manuscript.
S. M. Koksbang also acknowledges useful correspondence with Mustapha Ishak regarding the work presented in \cite{dallas-light}.
\newline\indent
Parts of the numerical computations in this work have been done using computing resources from the Center for Scientific Computing Aarhus.

\appendix
\section{ODEs for the perturbed FLRW metric in the Newtonian gauge (Spherical coordinates)}\label{gauge-appendix_sph}
In this appendix the set of ODEs used to obtain $D_A(z)$ for perturbed FLRW metrics in the Newtonian gauge in spherical coordinates are given.
\newline\newline
The four geodesic equations are:
\begin{equation}\label{eq:kt_ng}
\begin{split}
c^2(1+2\psi)\dot k^t = -2c^2\dot{\psi}k^t + c^2\psi_{,t}(k^t)^2 + [\psi_{,t}a^2 -\\ 
(1-2\psi)aa_{,t}][(k^r)^2 + r^2(k^{\theta})^2 + r^2\sin^2(\theta)(k^{\phi})^2]
\end{split}
\end{equation}
\begin{equation}\label{eq:kr_ng}
\begin{split}
(1-2\psi)a^2\dot k^r = 2a^2\dot{\psi}k^r -\\ 2(1-2\psi)aa_{,t}k^tk^r - c^2\psi_{,r}(k^t)^2 - a^2\psi_{,r}(k^r)^2 +\\ 
a^2r[(1-2\psi)-r\psi_{,r}]
[(k^{\theta})^2 + \sin^2(\theta)(k^{\phi})^2]
\end{split}
\end{equation}
\begin{equation}\label{eq:ktheta_ng}
\begin{split}
(1-2\psi)a^2r^2\dot k^{\theta} = 2\dot{\psi} a^2r^2k^{\theta} - 2(1-2\psi)ark^{\theta}(ra_{,t}k^t + ak^r) -\\ c^2\psi_{,\theta}(k^t)^2-
a^2\psi_{,\theta}[(k^r)^2 + r^2(k^{\theta})^2 + r^2\sin^2(\theta)(k^{\phi})^2] +\\ (1-2\psi)a^2r^2\sin(\theta)\cos(\theta)(k^{\phi})^2
\end{split}
\end{equation}
\begin{equation}\label{eq:kphi_ng}
\begin{split}
(1-2\psi)a^2r^2\sin^2(\theta)\dot k^{\phi} = 2\dot\psi a^2r^2\sin^2(\theta)k^{\phi} -\\ 2(1-2\psi) k^{\phi}ar\sin(\theta)
[a_{,t}k^tr\sin(\theta) + a\sin(\theta)k^r +
 ar\cos(\theta)k^{\theta}] -\\ 
c^2\psi_{,\phi}(k^t)^2 - \psi_{,\phi}a^2[(k^r)^2+r^2(k^{\theta})^2+r^2\sin^2(\theta)(k^{\phi})^2]
\end{split}
\end{equation}
Taking the partial temporal derivative of equation (\ref{eq:kt_ng}) one obtains:
\begin{equation}
\begin{split}
c^2(1+2\psi)\frac{d}{d\lambda}k^t_{,t} = -c^2(1+2\psi)k^t_{,\beta}k^{\beta}_{,t} -\\ 
2c^2 \psi_{,t}\dot k^t - 2c^2k^t(\psi_{,t\beta}k^{\beta} + \psi_{,\beta}k^{\beta}_{,t}) - 2c^2\dot\psi k^t_{,t} +\\ c^2\psi_{,tt}(k^t)^2 + 2c^2\psi_{,t} k^tk^t_{,t} + 2[\psi_{,t}a^2 - (1-2\psi)aa_{,t}][k^rk^r_{,t} +\\ 
r^2k^{\theta}k^{\theta}_{,t} + r^2\sin^2(\theta)k^{\phi}k^{\phi}_{,t}] +[\psi_{,tt}a^2 + 4\psi_{,t}aa_{,t} -\\ 
(1-2\psi)(a_{,t}^2+ aa_{,tt})][(k^r)^2 + r^2(k^{\theta})^2 + r^2\sin^2(\theta)(k^{\phi})^2]
\end{split}
\end{equation}
The partial $r, \theta$ and $\phi$ derivatives of \ref{eq:kt_ng} corresponds to the following three equations:
\begin{equation}
\begin{split}
c^2(1+2\psi)\frac{d}{d\lambda}k^t_{,r} = -c^2(1+2\psi)k^t_{,\beta}k^{\beta}_{,r}-2c^2\psi_{,r}\dot k^t -\\ 2c^2k^t[\psi_{,r\beta}k^{\beta} + \psi_{,\beta}k^\beta_{,r}]-
2c^2\dot \psi k^t_{,r} + c^2\psi_{,tr}(k^t)^2 +\\ 2c^2\psi_{,t}k^t_{,r}k^t + 2[\psi_{,t}a^2 - (1-2\psi)aa_{,t}]
[k^rk^r_{,r} + r(k^{\theta})^2 +\\ 
r^2k^{\theta}k^{\theta}_{,r} + r\sin^2(\theta)(k^{\phi})^2 + r^2\sin^2(\theta)k^{\phi}k^{\phi}_{,r}] + [\psi_{,tr}a^2 +\\ 2\psi_{,r}aa_{,t}][(k^r)^2 + r^2(k^{\theta})^2 + r^2\sin^2(\theta)(k^{\phi})^2]
\end{split}
\end{equation}
\begin{equation}
\begin{split}
c^2(1+2\psi)\frac{d}{d\lambda}k^t_{,\theta} = -c^2(1+2\psi)k^t_{,\beta}k^{\beta}_{,\theta} - 2c^2 \psi_{,\theta} \dot k^t -\\ 
2c^2k^t[\psi_{,\alpha\theta}k^{\alpha} + \psi_{,\alpha}k^{\alpha}_{,\theta}] -
2c^2\dot\psi k^t_{,\theta} + c^2\psi_{,t\theta}(k^t)^2+\\
2c^2\psi_{,t}k^tk^t_{,\theta} + 2[\psi_{,t}a^2 - (1-2\psi)aa_{,t}] 
[k^rk^r_{,\theta} + r^2k^{\theta}k^{\theta}_{,\theta} +\\ r^2\sin(\theta)\cos(\theta)(k^{\phi})^2 + r^2\sin^2(\theta)k^{\phi}k^{\phi}_{,\theta}] + [\psi_{,t\theta}a^2 +\\ 
2\psi_{,\theta}aa_{,t}]
[(k^r)^2 + r^2(k^{\theta})^2 + r^2\sin^2(\theta)(k^{\phi})^2]
\end{split}
\end{equation}
\begin{equation}
\begin{split}
c^2(1+2\psi)\frac{d}{d\lambda}k^t_{,\phi} = -c^2(1+2\psi)k^t_{,\beta}k^{\beta}_{,\phi} - 2c^2\psi_{,\phi}\dot k^t  -\\
2c^2k^t[\psi_{,\alpha\phi}k^{\alpha} + \psi_{,\alpha}k^{\alpha}_{,\phi}]-2c^2\dot{\psi} k^t_{,\phi}+c^2\psi_{,t\phi}(k^t)^2 +\\ 2c^2\psi_{,t}k^tk^t_{,\phi} + 2[\psi_{,t}a^2 - (1-2\psi)aa_{,t}][k^rk^r_{,\phi} +\\ r^2k^{\theta}k^{\theta}_{,\phi} +
r^2\sin^2(\theta)k^{\phi}k^{\phi}_{,\phi}] +
[\psi_{,t\phi} a^2 +\\ 
2\psi_{,\phi}aa_{,t}][(k^r)^2 +r^2(k^{\theta})^2 + r^2\sin^2(\theta) (k^{\phi})^2]
\end{split}
\end{equation}
\newline
The partial derivatives of equation (\ref{eq:kr_ng}) are:
\begin{equation}
\begin{split}
(1-2\psi)a^2\frac{d}{d\lambda}k^r_{,t} = -[2(1-2\psi)aa_{,t} - 2\psi_{,t}a^2]\dot k^r -\\ 
(1-2\psi)a^2k^r_{,\beta}k^{\beta}_{,t}
+2[2aa_{,t}\dot{\psi} + a^2(\psi_{,\beta}k^{\beta}_{,t} +\\ \psi_{,t\beta}k^{\beta})]k^r +
2a^2\dot\psi k^r_{,t}-2[(1-2\psi)(a_{,tt}a +\\ 
a_{,t}^2) - 2\psi_{,t}aa_{,t}]k^tk^r - 
2(1-2\psi)aa_{,t}(k^t_{,t}k^r +\\ 
k^tk^r_{,t}) - c^2\psi_{,tr}(k^t)^2-
2c^2\psi_{,r}k^tk^t_{,t} -[2aa_{,t}\psi_{,r} +\\ a^2\psi_{,tr}](k^r)^2 - 2a^2\psi_{,r}k^rk^r_{,t}
+2aa_{,t}r[(1-2\psi)-r\psi_{,r}][(k^{\theta})^2 +\\ \sin^2(\theta)(k^{\phi})^2] +
2a^2r[(1-2\psi)-r\psi_{,r}][k^{\theta}k^{\theta}_{,t} +\\ \sin^2(\theta)k^{\phi}k^{\phi}_{,t}] + a^2r[-2\psi_{,t}-r\psi_{,tr}][(k^{\theta})^2 + \sin^2(\theta)(k^{\phi})^2]
\end{split}
\end{equation}
\begin{equation}
\begin{split}
(1-2\psi)a^2\frac{d}{d\lambda}k^r_{,r} = -(1-2\psi)a^2k^r_{,\beta}k^{\beta}_{,r} + 2\psi_{,r}a^2\dot k^r +\\
2a^2\dot\psi k^r_{,r}+2a^2k^r[\psi_{,r\beta}k^{\beta} + \psi_{,\beta}k^{\beta}_{,r}] +
4\psi_{,r}aa_{,t}k^tk^r -\\
2(1-2\psi)aa_{,t}[k^t_{,r}k^r + k^tk^r_{,r}]-
c^2\psi_{,rr}(k^t)^2-2c^2\psi_{,r}k^tk^t_{,r} -\\ a^2\psi_{,rr}(k^r)^2 - 2a^2\psi_{,r}k^rk^r_{,r}+
a^2[(1-2\psi)-r\psi_{,r}][(k^{\theta})^2 +\\ \sin^2(\theta)(k^{\phi})^2] +a^2r[-3\psi_{,r} -r\psi_{,rr}]
[(k^{\theta})^2 +\\ 
\sin^2(\theta)(k^{\phi})^2]
+2ra^2[(1-2\psi)-r\psi_{,r}][k^{\theta}k^{\theta}_{,r} + \sin^2(\theta)k^{\phi}k^{\phi}_{,r}]
\end{split}
\end{equation}
\begin{equation}
\begin{split}
(1-2\psi)a^2\frac{d}{d\lambda}k^r_{,\theta} = -(1-2\psi)a^2k^r_{,\beta}k^{\beta}_{,\theta} +\\
2a^2\psi_{,\theta}\dot k^r + 2a^2k^r[\psi_{,\beta\theta}k^{\beta} + \psi_{,\beta}k^{\beta}_{,\theta}] +2a^2\dot\psi k^r_{,\theta}-\\
2(1-2\psi)aa_{,t}[k^t_{,\theta}k^r + k^tk^r_{,\theta}] + 4\psi_{,\theta}aa_{,t}k^tk^r -\\
c^2\psi_{,r\theta}(k^t)^2 - 2c^2\psi_{,r}k^tk^t_{,\theta} - a^2 \psi_{,r\theta}(k^r)^2 - 2a^2\psi_{,r}k^rk^r_{,\theta} +\\ 
a^2r[-2\psi_{,\theta} - r\psi_{,r\theta}][(k^{\theta})^2 + \sin^2(\theta)(k^{\phi})^2]
+2a^2r[(1-2\psi)-\\
r\psi_{,r}][k^{\theta}k^{\theta}_{,\theta} + \sin(\theta)\cos(\theta)(k^{\phi})^2 + \sin^2(\theta)k^{\phi}k^{\phi}_{,\theta}]
\end{split}
\end{equation}
\begin{equation}
\begin{split}
(1-2\psi)a^2\frac{d}{d\lambda}k^r_{,\phi} = -(1-2\psi)a^2k^r_{,\beta}k^{\beta}_{,\phi} +\\ 2a^2\psi_{,\phi}\dot k^r + 2a^2k^r[\psi_{,\beta\phi}k^{\beta} + \psi_{,\beta}k^{\beta}_{,\phi}] +2a^2\dot \psi k^r_{,\phi}-\\
2(1-2\psi)aa_{,t}(k^t_{,\phi}k^r + k^tk^r_{,\phi}) + 4\psi_{,\phi}aa_{,t}k^tk^r -\\ c^2\psi_{,r\phi}(k^t)^2-2c^2\psi_{,r}k^tk^t_{,\phi} - a^2\psi_{,r\phi}(k^r)^2-2a^2\psi_{,r}k^rk^r_{,\phi}+\\ a^2r[-2\psi_{,\phi}-r\psi_{,r\phi}][(k^{\theta})^2 + \sin^2(\theta)(k^{\phi})^2] +\\ 
2a^2r[(1-2\psi) - r\psi_{,r}][k^{\theta}k^{\theta}_{,\phi}
+\sin^2(\theta)k^{\phi}k^{\phi}_{,\phi}]
\end{split}
\end{equation}
\newline
The partial derivatives of equation (\ref{eq:ktheta_ng}) are:
\begin{equation}
\begin{split}
(1-2\psi)a^2r^2\frac{d}{d\lambda}k^{\theta}_{,t} = -(1-2\psi)a^2r^2k^{\theta}_{,\beta}k^{\beta}_{,t} -\\
r^2\dot k^{\theta}[-2\psi_{,t}a^2 + 2(1-2\psi)aa_{,t}] + 2r^2k^{\theta}[2aa_{,t}\dot{\psi}+\\
a^2(\psi_{,\beta t} k^{\beta} + \psi_{,\beta}k^{\beta}_{,t})] +2a^2r^2\dot\psi k^{\theta}_{,t}-2[(1-2\psi)a_{,t}rk^{\theta} -\\ 2ar\psi_{,t}k^{\theta} + (1-2\psi)ark^{\theta}_{,t}][ra_{,t}k^t + ak^r]-\\
2(1-2\psi)ark^{\theta}[ra_{,tt}k^t + ra_{,t}k^t_{,t} + a_{,t}k^r + ak^r_{,t}] -\\
c^2\psi_{,t\theta}(k^t)^2 -2c^2\psi_{,\theta}k^tk^t_{,t} -
[2aa_{,t}\psi_{,\theta} + a^2\psi_{,t\theta}][(k^r)^2 +\\ r^2(k^{\theta})^2 + r^2\sin^2(\theta)(k^{\phi})^2]-2a^2\psi_{,\theta}[k^rk^r_{,t} + r^2k^{\theta}k^{\theta}_{,t} +\\ r^2\sin^2(\theta)k^{\phi}k^{\phi}_{,t}]
+2r^2\sin(\theta)\cos(\theta)[(1-2\psi)aa_{,t}(k^{\phi})^2 -\\ \psi_{,t}a^2(k^{\phi})^2 + (1-2\psi)a^2k^{\phi}k^{\phi}_{,t}]
\end{split}
\end{equation}
\begin{equation}
\begin{split}
(1-2\psi)a^2r^2\frac{d}{d\lambda}k^{\theta}_{,r} = -(1-2\psi)a^2r^2k^{\theta}_{,\beta}k^{\beta}_{,r} -\\
a^2\dot k^{\theta}[2r(1-2\psi) - 2r^2\psi_{,r}] +2a^2k^{\theta}[2r\dot{\psi} +\\ 
r^2(\psi_{,r\beta}k^{\beta} + \psi_{,\beta}k^{\beta}_{,r})]+ 2a^2r^2\dot{\psi}k^{\theta}_{,r} -\\
2a[k^{\theta}(1-2\psi) -2r\psi_{,r}k^{\theta}+(1-2\psi)rk^{\theta}_{,r}][a_{,t}rk^t + ak^r]-\\
2(1-2\psi)ark^{\theta}[a_{,t}k^t + ra_{,t}k^t_{,r} + ak^r_{,r}] -c^2\psi_{,\theta r} (k^t)^2-\\
2c^2\psi_{,\theta}k^tk^t_{,r} - a^2\psi_{,\theta r}[(k^r)^2 +r^2(k^{\theta})^2 +r^2\sin^2(\theta)(k^{\phi})^2] -\\
2a^2\psi_{,\theta}[k^rk^r_{,r} + r(k^{\theta})^2 + r^2k^{\theta}k^{\theta}_{,r}+\\
r\sin^2(\theta)(k^{\phi})^2 + r^2\sin^2(\theta)k^{\phi}k^{\phi}_{,r}] + 2a^2\cos(\theta)\sin(\theta)\\
[r(1-2\psi)(k^{\phi})^2 -r^2\psi_{,r}(k^{\phi})^2 + (1-2\psi)r^2k^{\phi}k^{\phi}_{,r}]
\end{split}
\end{equation}
\begin{equation}
\begin{split}
(1-2\psi)a^2r^2\frac{d}{d\lambda}k^{\theta}_{,\theta} = -(1-2\psi)a^2r^2k^{\theta}_{,\beta}k^{\beta}_{,\theta}+\\
2\psi_{,\theta}a^2r^2\dot k^{\theta} + 2a^2r^2k^{\theta}[\psi_{,\beta\theta}k^{\beta} + \psi_{,\beta}k^{\beta}_{,\theta}] +\\ 
2\dot \psi a^2r^2k^{\theta}_{,\theta} -2ar[(1-2\psi)k^{\theta}_{,\theta} - 2\psi_{,\theta}k^{\theta}][ra_{,t}k^t + ak^r] -\\ 2ar(1-2\psi)k^{\theta}[ra_{,t}k^t_{,\theta}+ ak^r_{,\theta}]-
c^2\psi_{,\theta\theta}(k^t)^2 -\\ 2c^2\psi_{,\theta}k^tk^t_{,\theta} - a^2\psi_{,\theta\theta}[(k^r)^2 + r^2(k^{\theta})^2 +\\ r^2\sin^2(\theta)(k^{\phi})^2] - 2a^2\psi_{,\theta}[k^rk^r_{,\theta} + 
r^2k^{\theta}k^{\theta}_{,\theta} +\\ r^2\sin(\theta)\cos(\theta)(k^{\phi})^2 + r^2\sin^2(\theta)k^{\phi}k^{\phi}_{,\theta}] +\\ 
a^2r^2(k^{\phi})^2
[(1-2\psi)(\cos^2(\theta)-\sin^2(\theta)) -
2\psi_{,\theta}\sin(\theta)\cos(\theta)] +\\ 2(1-2\psi)a^2r^2\sin(\theta)\cos(\theta)k^{\phi}k^{\phi}_{,\theta}
\end{split}
\end{equation}
\begin{equation}
\begin{split}
(1-2\psi)a^2r^2\frac{d}{d\lambda}k^{\theta}_{,\phi} = -(1-2\psi)a^2r^2k^{\theta}_{,\beta}k^{\beta}_{,\phi} +\\
2a^2r^2\psi_{,\phi}\dot k^{\theta} + 2a^2r^2\dot \psi k^{\theta}_{,\phi} +
2a^2r^2k^{\theta}[\psi_{,\beta\phi}k^{\beta} + \psi_{,\beta}k^{\beta}_{,\phi}] -\\ 2[(1-2\psi)ark^{\theta}_{,\phi} - 2ar\psi_{,\phi}k^{\theta}][ra_{,t}k^t + ak^{r}] -\\ 2ar(1-2\psi)k^{\theta}[ra_{,t}k^t_{,\phi} + ak^r_{,\phi}]-
c^2\psi_{,\theta\phi}(k^t)^2 -\\ 
2c^2\psi_{,\theta}k^tk^t_{,\phi} - a^2\psi_{,\theta\phi}[(k^r)^2 + r^2(k^{\theta})^2 + r^2\sin^2(\theta)(k^{\phi})^2] -\\
2a^2\psi_{,\theta}[k^rk^r_{,\phi} + r^2k^{\theta}k^{\theta}_{,\phi} + r^2\sin^2(\theta)k^{\phi}k^{\phi}_{,\phi}]-\\
2a^2r^2\psi_{,\phi}\sin(\theta)\cos(\theta)(k^{\phi})^2 + 2a^2r^2(1-2\psi)\sin(\theta)\cos(\theta)k^{\phi}k^{\phi}_{,\phi}
\end{split}
\end{equation}
\newline
The four partial derivatives of equation (\ref{eq:kphi_ng}) are:
\begin{equation}
\begin{split}
(1-2\psi)a^2r^2\sin^2(\theta)\frac{d}{d\lambda}k^{\phi}_{,t} =\\
-(1-2\psi)a^2r^2\sin^2(\theta)k^{\phi}_{,\beta}k^{\beta}_{,t} -\\
2r^2\sin^2(\theta)a\dot k^{\phi}[-\psi_{,t}a + (1-2\psi)a_{,t}] +\\
2r^2\sin^2(\theta)[a^2k^{\phi}(\psi_{,\alpha t}k^{\alpha} +\\
\psi_{,\alpha}k^{\alpha}_{,t}) +2aa_{,t}k^{\phi}\dot\psi + a^2\dot\psi k^{\phi}_{,t}] -\\
2r\sin(\theta)[(1-2\psi)a_{,t}k^{\phi} +(1-2\psi)ak^{\phi}_{,t} -\\
2\psi_{,t}ak^{\phi}][a_{,t}k^tr\sin(\theta) + a\sin(\theta)k^r +\\
ar\cos(\theta)k^{\theta}]-2(1-2\psi)ar\sin(\theta)k^{\phi}[a_{,tt}k^tr\sin(\theta) +\\
a_{,t}k^t_{,t}r\sin(\theta) + a_{,t}\sin(\theta)k^r + a\sin(\theta)k^r_{,t}+\\
a_{,t} r\cos(\theta)k^{\theta} +ar\cos(\theta)k^{\theta}_{,t}]-c^2\psi_{,\phi t}(k^t)^2 -\\
2c^2\psi_{,\phi}k^tk^t_{,t} -[\psi_{,t\phi}a^2 +2aa_{,t}\psi_{,\phi}][(k^r)^2 +\\
r^2(k^{\theta})^2 +r^2\sin^2(\theta)(k^{\phi})^2] - 2a^2\psi_{,\phi}[k^rk^r_{,t} +\\
r^2k^{\theta}k^{\theta}_{,t} + r^2\sin^2(\theta)k^{\phi}k^{\phi}_{,t}]
\end{split}
\end{equation}
\begin{equation}
\begin{split}
a^2r^2\sin^2(\theta)(1-2\psi)\frac{d}{d\lambda}k^{\phi}_{,r} =\\
-a^2r^2\sin^2(\theta)(1-2\psi)k^{\phi}_{,\beta}k^{\beta}_{,r} - \\
2a^2r\sin^2(\theta)\dot k^{\phi}[1-2\psi -\psi_{,r}r] +\\
2a^2\sin^2(\theta)k^{\phi}[2r\dot \psi + r^2(\psi_{,r\beta}k^{\beta} +\\
\psi_{,\beta}k^{\beta}_{,r})]+2\dot \psi a^2r^2\sin^2(\theta)k^{\phi}_{,r} -\\
2a\sin(\theta)[(1-2\psi)k^{\phi} -2\psi_{,r}rk^{\phi} +\\
(1-2\psi)rk^{\phi}_{,r}][a_{,t}k^tr\sin(\theta) +a\sin(\theta)k^r +\\
ar\cos(\theta)k^{\theta}] - 2(1-2\psi)k^{\phi}ar\sin(\theta)\\
[a_{,t}k^t_{,r}r\sin(\theta)+a_{,t}k^t\sin(\theta) + a\sin(\theta)k^r_{,r} +\\ 
a\cos(\theta)k^{\theta} +ar\cos(\theta)k^{\theta}_{,r}] -\\ 
c^2\psi_{,\phi r}(k^t)^2-2c^2\psi_{,\phi}k^tk^t_{,r} - \psi_{,r\phi}a^2[(k^r)^2 +\\
r^2(k^{\theta})^2 +\\ r^2\sin^2(\theta)(k^{\phi})^2]-\\
2a^2\psi_{,\phi}[k^rk^r_{,r} + r(k^{\theta})^2 +r^2k^{\theta}k^{\theta}_{,r} +\\
r\sin^2(\theta)(k^{\phi})^2 + r^2\sin^2(\theta)k^{\phi}k^{\phi}_{,r}]
\end{split}
\end{equation}
\begin{equation}
\begin{split}
(1-2\psi)a^2r^2\sin^2(\theta)\frac{d}{d\lambda}k^{\phi}_{,\theta} =\\ -(1-2\psi)a^2r^2\sin^2(\theta)k^{\phi}_{,\beta}k^{\beta}_{,\theta}
-2a^2r^2\sin(\theta)\dot k^{\phi}\\
[-\psi_{,\theta}\sin(\theta) + (1-2\psi)\cos(\theta)] +
2a^2r^2k^{\phi}\\
[2\dot \psi \sin(\theta)\cos(\theta) + \sin^2(\theta)(\psi_{,\beta\theta}k^{\beta} + \psi_{,\beta}k^{\beta}_{,\theta})] +\\ 
2\dot\psi a^2r^2\sin^2(\theta)k^{\phi}_{,\theta}-
2ar[(1-2\psi)\sin(\theta)k^{\phi}_{,\theta} +\\ (1-2\psi)k^{\phi}\cos(\theta) - 2\psi_{,\theta}k^{\phi}\sin(\theta)]
[a_{,t}k^tr\sin(\theta) +\\ 
a\sin(\theta)k^r + ar\cos(\theta)k^{\theta}] -
2(1-2\psi)k^{\phi}ar\sin(\theta)\\
[a_{,t} k^t_{,\theta} \sin(\theta)r +
a_{,t}k^t\cos(\theta)r + a\cos(\theta)k^r + a\sin(\theta)k^r_{,\theta} +\\
ar\cos(\theta)k^{\theta}_{,\theta} - ar\sin(\theta)k^{\theta}]-
c^2\psi_{,\phi\theta}(k^t)^2-2c^2\psi_{,\phi}k^tk^t_{,\theta} -\\ \psi_{,\phi\theta}a^2[(k^r)^2 + r^2(k^{\theta})^2 +\\ r^2\sin^2(\theta)(k^{\phi})^2]
-2a^2\psi_{,\phi}[k^rk^r_{,\theta} + 
r^2k^{\theta}k^{\theta}_{,\theta} +\\ r^2\sin^2(\theta)k^{\phi}k^{\phi}_{,\theta} + r^2\sin(\theta)\cos(\theta)(k^{\phi})^2]
\end{split}
\end{equation}
\begin{equation}
\begin{split}
(1-2\psi)a^2r^2\sin^2(\theta)\frac{d}{d\lambda}k^{\phi}_{,\phi} =\\
-(1-2\psi)a^2r^2\sin^2(\theta)k^{\phi}_{,\beta}k^{\beta}_{,\phi}+\\
2a^2r^2\sin^2(\theta)\psi_{,\phi}\dot k^{\phi} +
2a^2r^2\sin^2 (\theta)\dot{\psi} k^{\phi}_{,\phi} +\\ 2a^2r^2\sin^2(\theta)k^{\phi}[\psi_{,\beta\phi}k^{\beta} + \psi_{,\beta}k^{\beta}_{,\phi}]-\\ 2ar\sin(\theta)[(1-2\psi)k^{\phi}_{,\phi} - 2\psi_{,\phi}k^{\phi}]\\
[a_{,t}k^tr\sin(\theta) + a\sin(\theta)k^r + ar\cos(\theta)k^{\theta}]-\\
2(1-2\psi)k^{\phi}ar\sin(\theta)[a_{,t}r\sin(\theta)k^t_{,\phi} +\\ 
a\sin(\theta)k^r_{,\phi} + ar\cos(\theta)k^{\theta}_{,\phi}]-
c^2\psi_{,\phi\phi}(k^t)^2-\\
2c^2\psi_{,\phi}k^tk^t_{,\phi} - \psi_{,\phi\phi}a^2[(k^r)^2 + r^2(k^{\theta})^2 +\\ 
r^2\sin^2(\theta)(k^{\phi})^2]-
2a^2\psi_{,\phi}[k^rk^r_{,\phi} + r^2k^{\theta}k^{\theta}_{,\phi} +\\ r^2\sin^2(\theta)k^{\phi}k^{\phi}_{,\phi}]
\end{split}
\end{equation}
The Christoffel symbols needed to obtain $D_A$ are:
\begin{equation}
\begin{split}
\Gamma_{tt}^t = \frac{\psi_{,t}}{1+2\psi}\\
\Gamma_{rt}^t = \frac{\psi_{,r}}{1+2\psi} \\
\Gamma_{\theta t}^t = \frac{\psi_{,\theta}}{1+2\psi}\\
\Gamma_{\phi t}^t = \frac{\psi_{,\phi}}{1+2\psi}\\
\Gamma_{tr}^r = \Gamma_{t\theta}^{\theta} = \Gamma_{t\phi}^{\phi} =  \frac{a\psi_{,t} + a_{,t}(2\psi-1)}{a(2\psi-1)}\\
\Gamma_{rr}^r = \frac{\psi_{,r}}{2\psi-1}\\
\Gamma_{r\theta}^{\theta} = \Gamma_{r\phi}^{\phi}= \frac{\psi_{,r}r + 2\psi-1}{r(2\psi-1)}\\
\Gamma_{\theta r}^r =\Gamma_{\theta\theta}^{\theta} =\frac{\psi_{,\theta}}{2\psi-1} \\
\Gamma_{\theta \phi}^{\phi} = \frac{\psi_{,\theta}\sin(\theta) + (2\psi-1)\cos(\theta)}{(2\psi-1)\sin(\theta)}\\
\Gamma_{\phi r}^r = \Gamma_{\phi\theta}^{\theta} = \Gamma_{\phi\phi}^{\phi} = \frac{\psi_{,\phi}}{2\psi-1}
\end{split}
\end{equation}
Using the Christoffel symbols, the following differential equation for the angular diameter distance is obtained:
\begin{equation}
\begin{split}
\frac{d}{d\lambda}D_A = \frac{1}{2}D_A( k^t_{,t} +k^r_{,r} + k^{\theta}_{,\theta} + k^{\phi}_{,\phi} +\\ 
2\dot{\psi}\frac{4\psi + 1}{4\psi^2 -1} 
+3\frac{a_{,t}}{a}k^t + \frac{2}{r}k^r + \frac{\cos(\theta)}{\sin(\theta)}k^{\theta}) 
\end{split}
\end{equation}
\newline\newline
When setting the initial conditions, the null-condition and its partial derivatives are used to ensure that the geodesic will be null. the null-condition and its derivatives are also used for checking the accuracy of the numerical computations. The null-condition is:
\begin{equation}
\begin{split}
k^{\alpha}k_{\alpha} = k^{\alpha}k^{\beta}g_{\alpha\beta} = -c^2(1+2\psi)(k^t)^2+\\
(1-2\psi)a^2[(k^r)^2 + r^2(k^{\theta})^2 + r^2\sin^2(\theta)(k^{\phi})^2] = 0
\end{split}
\end{equation}
The four partial derivatives are:
\begin{equation}
\begin{split}
-2c^2\psi_{,t}(k^t)^2-2c^2(1+2\psi)k^tk^t_{,t} + 2[(1-2\psi)aa_{,t}-\\
\psi_{,t}a^2][(k^r)^2 + r^2(k^{\theta})^2 + r^2\sin^2(\theta)(k^{\phi})^2]+\\
2(1-2\psi)a^2[k^rk^r_{,t} +
r^2k^{\theta}k^{\theta}_{,t} + r^2\sin^2(\theta)k^{\phi}k^{\phi}_{,t}] = 0
\end{split}
\end{equation}
\begin{equation}
\begin{split}
-2c^2\psi_{,r}(k^t)^2 -2c^2(1+2\psi)k^tk^t_{,r}-2\psi_{,r}a^2[(k^r)^2 +\\ r^2(k^{\theta})^2 + r^2\sin^2(\theta)(k^{\phi})^2]+
2(1-2\psi)a^2[k^rk^r_{,r} + r(k^{\theta})^2 +\\ r^2k^{\theta}k^{\theta}_{,r} + r\sin^2(\theta)(k^{\phi})^2 + r^2\sin^2(\theta)k^{\phi}k^{\phi}_{,r}]=0
\end{split}
\end{equation}
\begin{equation}
\begin{split}
-2c^2\psi_{,\theta}(k^t)^2 - 2c^2(1+2\psi)k^tk^t_{,\theta}-2\psi_{,\theta}a^2[(k^r)^2 +\\ r^2(k^{\theta})^2 + r^2\sin^2(\theta)(k^{\phi})^2]+
2(1-2\psi)a^2(k^rk^r_{,\theta} + r^2k^{\theta}k^{\theta}_{,\theta} +\\
 r^2\sin^2(\theta)k^{\phi}k^{\phi}_{,\theta} + r^2\cos(\theta)\sin(\theta)(k^{\phi})^2)=0
\end{split}
\end{equation}
\begin{equation}
\begin{split}
-2c^2\psi_{,\phi}(k^t)^2 - 2c^2(1+2\psi)k^tk^t_{,\phi} - 2\psi_{,\phi}a^2[(k^r)^2 +\\ r^2(k^{\theta})^2 + r^2\sin^2(\theta)(k^{\phi})^2]+
2(1-2\psi)a^2[k^rk^r_{,\phi} +\\
r^2k^{\theta}k^{\theta}_{,\phi} + r^2\sin^2(\theta)k^{\phi}k^{\phi}_{,\phi}] = 0
\end{split}
\end{equation}
\newline\newline
For an observer placed at the origin, the initial conditions must be in accordance with a radial null-geodesic. Aside from the trivial constraints this sets on the initial conditions, two constraints are worth mentioning. First, the initial condition of $k^t_{,t}$ should be determined from the following expression:
\begin{equation}
\begin{split}
k^t_{,t} = \left[  -\frac{1}{2}  \int_{t_0}^t\! 3\frac{T_{,t}}{T}+\frac{R_{,t}}{R} \right]_{,t}k^t =\\ -\frac{1}{2}\left[ 3\frac{T_{,t}}{T} + \frac{R_{,t}}{R}\right]k^t =  \left[ \frac{-3\psi_{,t}}{1+2\psi} + \frac{\psi_{,t}}{1-2\psi}-\frac{a_{,t}}{a}\right] k^t
\end{split}
\end{equation}
The initial condition of $k^r_{,r}$ is determined from the partial $r$-derivative of the null-condition with $k^t_{,r}|_{0} = -\frac{1}{2} \left( \int_{t_0}^t\! 3\frac{T_{,t}}{T}+\frac{R_{,t}}{R} dt\ \right)_{,r}|_{0}  = 0$, {\em i.e.} $k^r_{,r}|_0 = \frac{\psi_{,r}}{(1-2\psi)a^2k^r}(c^2(k^t)^2 + a^2(k^r)^2)$.
\newline\newline
See {\em e.g.} \cite{dallas-light} for details on how these initial conditions can be derived.

\section{ODEs for the quasi-spherical Szekeres model in spherical coordinates}\label{Szekeres-appendix}
In this appendix, the ODEs used to obtain $D_A(z)$ in quasi-spherical Szekeres models are given in expanded form (in spherical coordinates).
\newline\newline
The ODEs are solved by using a gsl ODE solver. In order to do this, it is necessary to eliminate  $\dot k^{\theta}$ and $\dot k^{\phi}$ from equation (\ref{eq:geod1}) and 
$\frac{d}{d\lambda}(k^{\theta}_{,\alpha})$ and $\frac{d}{d\lambda}(k^{\phi}_{,\alpha})$ from equation (\ref{eq:radius}). This yields two new equations which are given below.
\newline\indent
The first equation is obtained by using equations (\ref{eq:geod2}) and (\ref{eq:geod3}) to eliminate $\dot k^{\theta}$ and $\dot k^{\phi}$ in equation (\ref{eq:geod1}). The resulting equation is:
\begin{equation}
\begin{split}
\dot k^r +\frac{1}{R-\frac{\Phi^2}{P}-\frac{\Theta^2}{F}}[
\dot R k^r + \dot \Phi k^{\phi}-\\
\frac{\Phi}{P} \dot P k^{\phi} - \frac{\Phi}{P} \dot{\Phi} k^r+\frac{\Phi}{2P} R_{,\phi}(k^r)^2 +\\
\frac{\Phi}{P} \Theta_{,\phi} k^{\theta}k^r +
\frac{\Phi}{P} \Phi_{,\phi}k^{\phi}k^r + \dot \Theta k^{\theta} -\frac{\Theta}{F} \dot F k^{\theta} -\\ 
\frac{\Theta}{F} \dot \Theta k^r +\frac{\Theta}{2F} R_{,\theta} (k^r)^2 +\frac{\Theta}{2F}P_{,\theta}(k^{\phi})^2 +\\ \frac{\Theta}{F} \Theta_{,\theta} k^{\theta}k^r +
\frac{\Theta}{F} \Phi_{,\theta}k^{\phi}k^r -\\ \frac{1}{2}(R_{,r}(k^r)^2+F_{,r}(k^{\theta})^2+\\
P_{,r}(k^{\phi})^2+2\Theta_{,r}k^{\theta}k^r+2\Phi_{,r}k^{\phi}k^r)]=0
\end{split}
\end{equation}
Equivalently, equations (\ref{eq:theta}) and (\ref{eq:phi}) are used to eliminate $\frac{d}{d\lambda}(k^{\theta}_{,\alpha})$ and $\frac{d}{d\lambda}(k^{\phi}_{,\alpha})$ from equation (\ref{eq:radius}) to obtain:
\begin{equation}
\begin{split}
\frac{d}{d\lambda}(k^r_{,\alpha}) = -k^r_{,\beta}k^{\beta}_{,\alpha} -
\frac{1}{R-\frac{\Phi^2}{P}-\frac{\Theta^2}{F}}
(k^r_{,\alpha}(\dot R -\frac{\Phi}{P}\dot \Phi -\\ \frac{\Theta}{F}\dot \Theta)+ \dot k^r(R_{,\alpha} - \frac{\Phi}{P}\Phi_{,\alpha} - \frac{\Theta}{F}\Theta_{,\alpha}) + 
k^r(R_{,\alpha\beta}k^{\beta} +\\
R_{,\beta}k^{\beta}_{,\alpha}) + (\Phi_{,\alpha\beta}k^{\beta} + \Phi_{,\beta}k^{\beta}_{,\alpha})(k^{\phi}-\frac{\Phi}{P}k^r)  +\\ (\Theta_{,\alpha\beta}k^{\beta} +\Theta_{,\beta}k^{\beta}_{,\alpha})(k^{\theta} - \frac{\Theta}{F}k^r) +
\dot k^{\phi}(\Phi_{,\alpha} -\\ 
\frac{\Phi}{P}P_{,\alpha}) + k^{\phi}_{,\alpha}(\dot \Phi-\frac{\Phi}{P}\dot P) + 
\dot k^{\theta}(\Theta_{,\alpha} -
\frac{\Theta}{F}F_{,\alpha}) +\\
k^{\theta}_{,\alpha}(\dot \Theta -
\frac{\Theta}{F}\dot F)-(\frac{1}{2}R_{,r\alpha}(k^r)^2 + 
R_{,r}k^rk^r_{,\alpha} + 
\frac{1}{2}F_{,r\alpha}(k^{\theta})^2 +\\ F_{,r}k^{\theta}k^{\theta}_{,\alpha} + \frac{1}{2}P_{,r\alpha}(k^{\phi})^2 + P_{,r}k^{\phi}k^{\phi}_{,\alpha} +
\Theta_{,r\alpha}k^{\theta}k^r +\\ \Theta_{,r}(k^{\theta}_{,\alpha}k^r + k^{\theta}k^r_{,\alpha}) +
\Phi_{,r\alpha}k^{\phi}k^r + \Phi_{,r}(k^{\phi}_{,\alpha}k^r +\\ k^{\phi}k^r_{,\alpha})) - \frac{\Phi}{P}k^{\phi}(P_{,\alpha\beta}k^{\beta} + P_{,\beta}k^{\beta}_{,\alpha}) + \frac{\Phi}{P}(\frac{1}{2}R_{,\phi\alpha}(k^r)^2 +\\ R_{,\phi}k^rk^r_{,\alpha} +
\Theta_{,\alpha\phi}k^{\theta}k^r + 
\Theta_{,\phi}(k^{\theta}_{,\alpha}k^r + k^{\theta}k^r_{,\alpha}) + \Phi_{,\alpha\phi}k^{\phi}k^r +\\
\Phi_{,\phi}(k^{\phi}_{,\alpha}k^r + k^{\phi}k^r_{,\alpha})) - \frac{\Theta}{F}k^{\theta}(F_{,\alpha\beta}k^{\beta} +\\ F_{,\beta}k^{\beta}_{,\alpha})+\frac{\Theta}{F}(\frac{1}{2}R_{,\theta\alpha}(k^r)^2 +
R_{,\theta}k^rk^r_{,\alpha} + \frac{1}{2}P_{,\theta\alpha}(k^{\phi})^2 +\\ P_{,\theta}k^{\phi}k^{\phi}_{,\alpha} + \Theta_{,\alpha\theta}k^{\theta}k^r + \Theta_{,\theta}(k^{\theta}_{,\alpha}k^r + k^{\theta}k^r_{,\alpha}) +\\
\Phi_{,\theta\alpha}k^{\phi}k^r + \Phi_{,\theta}(k^{\phi}_{,\alpha}k^r + k^{\phi}k^r_{,\alpha})))
\end{split}
\end{equation}
\newline
Inserting $\alpha = t,r,\theta,\phi$ into equations (\ref{eq:time}), (\ref{eq:theta}), (\ref{eq:phi}) and the equation above, gives 16 ODEs for $k^{\beta}_{,\alpha}$. The four equations for $k^r_{,\alpha}$ above are (expanding the $\beta$-sums):
\begin{equation}
\begin{split}
\frac{d}{d\lambda}(k^r_{,t}) = -(k^r_{,t}k^{t}_{,t} + k^r_{,r}k^{r}_{,t} + k^r_{,\theta}k^{\theta}_{,t} + k^r_{,\phi}k^{\phi}_{,t}) -\\ \frac{1}{R-\frac{\Phi^2}{P}-\frac{\Theta^2}{F}}
(k^r_{,t}(\dot R -\frac{\Phi}{P}\dot \Phi - \frac{\Theta}{F}\dot \Theta)+\\ 
\dot k^r(R_{,t} - \frac{\Phi}{P}\Phi_{,t} - \frac{\Theta}{F}\Theta_{,t}) + 
k^r(R_{,tt}k^{t} + R_{,tr}k^{r} + R_{,t\theta}k^{\theta} +\\ R_{,t\phi}k^{\phi} + R_{,t}k^{t}_{,t}+
R_{,r}k^{r}_{,t}+ R_{,\theta}k^{\theta}_{,t}+ R_{,\phi}k^{\phi}_{,t}) +\\ 
(\Phi_{,tt}k^{t} + \Phi_{,tr}k^{r}+ \Phi_{,t\theta}k^{\theta}+ \Phi_{,t\phi}k^{\phi}+ \Phi_{,t}k^{t}_{,t} +\\ 
\Phi_{,r}k^{r}_{,t} + \Phi_{,\theta}k^{\theta}_{,t} + \Phi_{,\phi}k^{\phi}_{,t})(k^{\phi}-\frac{\Phi}{P}k^r)  +(\Theta_{,tt}k^{t} +\\
\Theta_{,tr}k^{r}+\Theta_{,t\theta}k^{\theta}+\Theta_{,t\phi}k^{\phi} + \Theta_{,t}k^{t}_{,t} +\\
\Theta_{,r}k^{r}_{,t} + \Theta_{,\theta}k^{\theta}_{,t} + \Theta_{,\phi}k^{\phi}_{,t})(k^{\theta} - \frac{\Theta}{F}k^r) + 
\dot k^{\phi}(\Phi_{,t} -\\
\frac{\Phi}{P}P_{,t}) + k^{\phi}_{,t}(\dot \Phi-\frac{\Phi}{P}\dot P) + \dot k^{\theta}(\Theta_{,t} -
\frac{\Theta}{F}F_{,t}) +\\ 
k^{\theta}_{,t}(\dot \Theta - \frac{\Theta}{F}\dot F)-(\frac{1}{2}R_{,rt}(k^r)^2 + 
R_{,r}k^rk^r_{,t} + \frac{1}{2}F_{,rt}(k^{\theta})^2 +\\ F_{,r}k^{\theta}k^{\theta}_{,t} + \frac{1}{2}P_{,rt}(k^{\phi})^2 + P_{,r}k^{\phi}k^{\phi}_{,t} + \Theta_{,rt}k^{\theta}k^r +\\ \Theta_{,r}(k^{\theta}_{,t}k^r + k^{\theta}k^r_{,t}) + \Phi_{,rt}k^{\phi}k^r + \Phi_{,r}(k^{\phi}_{,t}k^r + k^{\phi}k^r_{,t})) -\\
\frac{\Phi}{P}k^{\phi}(P_{,tt}k^{t} +P_{,tr}k^{r}+P_{,t\theta}k^{\theta}+ P_{,t}k^{t}_{,t} + P_{,r}k^{r}_{,t} + P_{,\theta}k^{\theta}_{,t}) +\\ \frac{\Phi}{P}(\frac{1}{2}R_{,\phi t}(k^r)^2 + R_{,\phi}k^rk^r_{,t} + \Theta_{,t\phi}k^{\theta}k^r + 
\Theta_{,\phi}(k^{\theta}_{,t}k^r +\\
k^{\theta}k^r_{,t}) + \Phi_{,t\phi}k^{\phi}k^r + \Phi_{,\phi}(k^{\phi}_{,t}k^r +\\ k^{\phi}k^r_{,t})) - \frac{\Theta}{F}k^{\theta}(F_{,tt}k^{t} + F_{,tr}k^{r} +\\ 
F_{,t}k^{t}_{,t} + F_{,r}k^{r}_{,t})+ \frac{\Theta}{F}(\frac{1}{2}R_{,\theta t}(k^r)^2 +\\
R_{,\theta}k^rk^r_{,t} + \frac{1}{2}P_{,\theta t}(k^{\phi})^2 + P_{,\theta}k^{\phi}k^{\phi}_{,t} + \Theta_{,t\theta}k^{\theta}k^r +\\ \Theta_{,\theta}(k^{\theta}_{,t}k^r + k^{\theta}k^r_{,t}) + \Phi_{,\theta t}k^{\phi}k^r + \Phi_{,\theta}(k^{\phi}_{,t}k^r + k^{\phi}k^r_{,t})))
\end{split}
\end{equation}
\begin{equation}
\begin{split}
\frac{d}{d\lambda}(k^r_{,r}) = -(k^r_{,t}k^{t}_{,r} + k^r_{,r}k^{r}_{,r} + k^r_{,\theta}k^{\theta}_{,r} + k^r_{,\phi}k^{\phi}_{,r}) -\\ \frac{1}{R-\frac{\Phi^2}{P}-\frac{\Theta^2}{F}}
(k^r_{,r}(\dot R -\frac{\Phi}{P}\dot \Phi - \frac{\Theta}{F}\dot \Theta)+\\ 
\dot k^r(R_{,r} - \frac{\Phi}{P}\Phi_{,r} - \frac{\Theta}{F}\Theta_{,r}) + 
k^r(R_{,rt}k^{t} + R_{,rr}k^{r} +\\
R_{,r\theta}k^{\theta} + R_{,r\phi}k^{\phi} + R_{,t}k^{t}_{,r}+ R_{,r}k^{r}_{,r}+ R_{,\theta}k^{\theta}_{,r}+\\ R_{,\phi}k^{\phi}_{,r}) + (\Phi_{,rt}k^{t} + \Phi_{,rr}k^{r}+ \Phi_{,r\theta}k^{\theta}+ \Phi_{,r\phi}k^{\phi}+\\ \Phi_{,t}k^{t}_{,r} + \Phi_{,r}k^{r}_{,r} + \Phi_{,\theta}k^{\theta}_{,r} + \Phi_{,\phi}k^{\phi}_{,r})(k^{\phi}-\frac{\Phi}{P}k^r)  +\\
(\Theta_{,rt}k^{t} + \Theta_{,rr}k^{r}+ \Theta_{,r\theta}k^{\theta}+\Theta_{,r\phi}k^{\phi} + \Theta_{,t}k^{t}_{,r} +\\
\Theta_{,r}k^{r}_{,r} + \Theta_{,\theta}k^{\theta}_{,r} + \Theta_{,\phi}k^{\phi}_{,r})(k^{\theta} - \frac{\Theta}{F}k^r) + 
\dot k^{\phi}(\Phi_{,r} -\\ 
\frac{\Phi}{P}P_{,r}) + k^{\phi}_{,r}(\dot \Phi-\frac{\Phi}{P}\dot P) + \dot k^{\theta}(\Theta_{,r} -
\frac{\Theta}{F}F_{,r}) +\\ 
k^{\theta}_{,r}(\dot \Theta - \frac{\Theta}{F}\dot F)-(\frac{1}{2}R_{,rr}(k^r)^2 + 
R_{,r}k^rk^r_{,r} + \frac{1}{2}F_{,rr}(k^{\theta})^2 +\\ F_{,r}k^{\theta}k^{\theta}_{,r} + \frac{1}{2}P_{,rr}(k^{\phi})^2 + P_{,r}k^{\phi}k^{\phi}_{,r} + \Theta_{,rr}k^{\theta}k^r + \Theta_{,r}(k^{\theta}_{,r}k^r +\\ 
k^{\theta}k^r_{,r}) + \Phi_{,rr}k^{\phi}k^r + \Phi_{,r}(k^{\phi}_{,r}k^r + k^{\phi}k^r_{,r})) -
\frac{\Phi}{P}k^{\phi}(P_{,rt}k^{t} +\\ P_{,rr}k^{r}+P_{,r\theta}k^{\theta}+ P_{,t}k^{t}_{,r} + P_{,r}k^{r}_{,r} + P_{,\theta}k^{\theta}_{,r}) +\\ \frac{\Phi}{P}(\frac{1}{2}R_{,\phi r}(k^r)^2 + R_{,\phi}k^rk^r_{,r} + \Theta_{,r\phi}k^{\theta}k^r + 
\Theta_{,\phi}(k^{\theta}_{,r}k^r +\\ 
k^{\theta}k^r_{,r}) + \Phi_{,r\phi}k^{\phi}k^r + \Phi_{,\phi}(k^{\phi}_{,r}k^r + k^{\phi}k^r_{,r})) - \frac{\Theta}{F}k^{\theta}(F_{,rt}k^{t} +\\
F_{,rr}k^{r} + F_{,t}k^{t}_{,r} + F_{,r}k^{r}_{,r})+\frac{\Theta}{F}(\frac{1}{2}R_{,\theta r}(k^r)^2 +R_{,\theta}k^rk^r_{,r} +\\
\frac{1}{2}P_{,\theta r}(k^{\phi})^2 + P_{,\theta}k^{\phi}k^{\phi}_{,r} + \Theta_{,r\theta}k^{\theta}k^r +\\ \Theta_{,\theta}(k^{\theta}_{,r}k^r + k^{\theta}k^r_{,r}) + \Phi_{,\theta r}k^{\phi}k^r + \Phi_{,\theta}(k^{\phi}_{,r}k^r + k^{\phi}k^r_{,r})))
\end{split}
\end{equation}
\begin{equation}
\begin{split}
\frac{d}{d\lambda}(k^r_{,\theta}) = -(k^r_{,t}k^{t}_{,\theta} + k^r_{,r}k^{r}_{,\theta} + k^r_{,\theta}k^{\theta}_{,\theta} + k^r_{,\phi}k^{\phi}_{,\theta}) -\\ \frac{1}{R-\frac{\Phi^2}{P}-\frac{\Theta^2}{F}}
(k^r_{,\theta}(\dot R -\frac{\Phi}{P}\dot \Phi -\\ \frac{\Theta}{F}\dot \Theta)+
\dot k^r(R_{,\theta} - \frac{\Phi}{P}\Phi_{,\theta} - \frac{\Theta}{F}\Theta_{,\theta}) + \\
k^r(R_{,\theta t}k^{t} + R_{,\theta r}k^{r} + R_{,\theta\theta}k^{\theta} + R_{,\theta\phi}k^{\phi} + R_{,t}k^{t}_{,\theta}+\\
R_{,r}k^{r}_{,\theta}+ R_{,\theta}k^{\theta}_{,\theta}+ R_{,\phi}k^{\phi}_{,\theta}) + (\Phi_{,\theta t}k^{t} + \Phi_{,\theta r}k^{r}+\\
\Phi_{,\theta\theta}k^{\theta}+ \Phi_{,\theta\phi}k^{\phi}+ \Phi_{,t}k^{t}_{,\theta} +
\Phi_{,r}k^{r}_{,\theta} +\\ 
\Phi_{,\theta}k^{\theta}_{,\theta} +\Phi_{,\phi}k^{\phi}_{,\theta})(k^{\phi}-\frac{\Phi}{P}k^r)  +\\
(\Theta_{,\theta t}k^{t} + \Theta_{,\theta r}k^{r}+\Theta_{,\theta\theta}k^{\theta}+\Theta_{,\theta\phi}k^{\phi} +\\ 
\Theta_{,t}k^{t}_{,\theta} + \Theta_{,r}k^{r}_{,\theta} + \Theta_{,\theta}k^{\theta}_{,\theta} + \Theta_{,\phi}k^{\phi}_{,\theta})(k^{\theta} - \frac{\Theta}{F}k^r) +\\ 
\dot k^{\phi}(\Phi_{,\theta} - \frac{\Phi}{P}P_{,\theta}) + k^{\phi}_{,\theta}(\dot \Phi-\frac{\Phi}{P}\dot P) + \dot k^{\theta}\Theta_{,\theta} +\\
k^{\theta}_{,\theta}(\dot \Theta - \frac{\Theta}{F}\dot F)-(\frac{1}{2}R_{,r\theta}(k^r)^2 + 
R_{,r}k^rk^r_{,\theta} + F_{,r}k^{\theta}k^{\theta}_{,\theta} +\\ \frac{1}{2}P_{,r\theta}(k^{\phi})^2 + P_{,r}k^{\phi}k^{\phi}_{,\theta} + \Theta_{,r\theta}k^{\theta}k^r + \Theta_{,r}(k^{\theta}_{,\theta}k^r + k^{\theta}k^r_{,\theta}) +\\ \Phi_{,r\theta}k^{\phi}k^r + \Phi_{,r}(k^{\phi}_{,\theta}k^r + k^{\phi}k^r_{,\theta})) -\frac{\Phi}{P}k^{\phi}(P_{,\theta t}k^{t} +P_{,\theta r}k^{r}+\\
P_{,\theta\theta}k^{\theta}+ P_{,t}k^{t}_{,\theta} + P_{,r}k^{r}_{,\theta} + P_{,\theta}k^{\theta}_{,\theta}) + \frac{\Phi}{P}(\frac{1}{2}R_{,\phi\theta}(k^r)^2 +\\ R_{,\phi}k^rk^r_{,\theta} + \Theta_{,\theta\phi}k^{\theta}k^r + 
\Theta_{,\phi}(k^{\theta}_{,\theta}k^r + k^{\theta}k^r_{,\theta}) + \Phi_{,\theta\phi}k^{\phi}k^r +\\ 
\Phi_{,\phi}(k^{\phi}_{,\theta}k^r + k^{\phi}k^r_{,\theta})) - \frac{\Theta}{F}k^{\theta}(F_{,t}k^{t}_{,\theta} + F_{,r}k^{r}_{,\theta})+\frac{\Theta}{F}(\frac{1}{2}R_{,\theta\theta}(k^r)^2 \\+R_{,\theta}k^rk^r_{,\theta} + \frac{1}{2}P_{,\theta\theta}(k^{\phi})^2 + P_{,\theta}k^{\phi}k^{\phi}_{,\theta} + \Theta_{,\theta\theta}k^{\theta}k^r +\\ \Theta_{,\theta}(k^{\theta}_{,\theta}k^r +
k^{\theta}k^r_{,\theta}) + 
\Phi_{,\theta\theta}k^{\phi}k^r + \Phi_{,\theta}(k^{\phi}_{,\theta}k^r + k^{\phi}k^r_{,\theta})))
\end{split}
\end{equation}
\begin{equation}
\begin{split}
\frac{d}{d\lambda}(k^r_{,\phi}) = -(k^r_{,t}k^{t}_{,\phi} + k^r_{,r}k^{r}_{,\phi} + k^r_{,\theta}k^{\theta}_{,\phi} + k^r_{,\phi}k^{\phi}_{,\phi}) -\\ \frac{1}{R-\frac{\Phi^2}{P}-\frac{\Theta^2}{F}}
(k^r_{,\phi}(\dot R -\frac{\Phi}{P}\dot \Phi - \frac{\Theta}{F}\dot \Theta)+\\ 
\dot k^r(R_{,\phi} - \frac{\Phi}{P}\Phi_{,\phi} - \frac{\Theta}{F}\Theta_{,\phi}) + 
k^r(R_{,\phi t}k^{t} + R_{,\phi r}k^{r} +\\
R_{,\phi\theta}k^{\theta} + R_{,\phi\phi}k^{\phi} + R_{,t}k^{t}_{,\phi}+ R_{,r}k^{r}_{,\phi}+ R_{,\theta}k^{\theta}_{,\phi}+\\ 
R_{,\phi}k^{\phi}_{,\phi}) + (\Phi_{,\phi t}k^{t} + \Phi_{,\phi r}k^{r}+ \Phi_{,\phi\theta}k^{\theta}+\\ 
\Phi_{,\phi\phi}k^{\phi}+ \Phi_{,t}k^{t}_{,\phi} + 
\Phi_{,r}k^{r}_{,\phi} + \Phi_{,\theta}k^{\theta}_{,\phi} + \Phi_{,\phi}k^{\phi}_{,\phi})\\
(k^{\phi}-\frac{\Phi}{P}k^r)  +(\Theta_{,\phi t}k^{t} + \Theta_{,\phi r}k^{r}+\Theta_{,\phi\theta}k^{\theta}+\\
\Theta_{,\phi\phi}k^{\phi} + \Theta_{,t}k^{t}_{,\phi} + \Theta_{,r}k^{r}_{,\phi} + \Theta_{,\theta}k^{\theta}_{,\phi} + \Theta_{,\phi}k^{\phi}_{,\phi})\\
(k^{\theta} - \frac{\Theta}{F}k^r) + 
\dot k^{\phi}\Phi_{,\phi} + k^{\phi}_{,\phi}(\dot \Phi-\frac{\Phi}{P}\dot P) +\\
\dot k^{\theta}\Theta_{,\phi} +
k^{\theta}_{,\phi}(\dot \Theta - \frac{\Theta}{F}\dot F)-(\frac{1}{2}R_{,r\phi}(k^r)^2 +\\ 
R_{,r}k^rk^r_{,\phi} + F_{,r}k^{\theta}k^{\theta}_{,\phi} + 
P_{,r}k^{\phi}k^{\phi}_{,\phi} + \Theta_{,r\phi}k^{\theta}k^r +\\ \Theta_{,r}(k^{\theta}_{,\phi}k^r + k^{\theta}k^r_{,\phi}) + \Phi_{,r\phi}k^{\phi}k^r + \Phi_{,r}(k^{\phi}_{,\phi}k^r + k^{\phi}k^r_{,\phi})) -\\
\frac{\Phi}{P}k^{\phi}(P_{,t}k^{t}_{,\phi} + P_{,r}k^{r}_{,\phi} + P_{,\theta}k^{\theta}_{,\phi}) + \frac{\Phi}{P}(\frac{1}{2}R_{,\phi\phi}(k^r)^2 +\\ R_{,\phi}k^rk^r_{,\phi} + \Theta_{,\phi\phi}k^{\theta}k^r + 
\Theta_{,\phi}(k^{\theta}_{,\phi}k^r + k^{\theta}k^r_{,\phi}) +\\ \Phi_{,\phi\phi}k^{\phi}k^r + \Phi_{,\phi}(k^{\phi}_{,\phi}k^r + k^{\phi}k^r_{,\phi})) - \frac{\Theta}{F}k^{\theta}(F_{,t}k^{t}_{,\phi} + F_{,r}k^{r}_{,\phi})\\
+\frac{\Theta}{F}(\frac{1}{2}R_{,\theta\phi}(k^r)^2 +R_{,\theta}k^rk^r_{,\phi} +\\
P_{,\theta}k^{\phi}k^{\phi}_{,\phi} + \Theta_{,\phi\theta}k^{\theta}k^r + \Theta_{,\theta}(k^{\theta}_{,\phi}k^r +\\ 
k^{\theta}k^r_{,\phi}) + \Phi_{,\theta\phi}k^{\phi}k^r + \Phi_{,\theta}(k^{\phi}_{,\phi}k^r + k^{\phi}k^r_{,\phi})))
\end{split}
\end{equation}
\newline
The four equations corresponding to equation (\ref{eq:time}) are:
\begin{equation}
\begin{split}
\frac{d}{d\lambda} (k^t_{,t}) = -[k^t_{,t} k^{t}_{,t} + k^t_{,r} k^{r}_{,t} +k^t_{,\theta} k^{\theta}_{,t} + k^t_{,\phi} k^{\phi}_{,t}] -\\
\frac{1}{c^2}[\frac{1}{2} R_{,tt} (k^r)^2+R_{,t} k^r k^r_{,t} + \frac{1}{2} F_{,tt} (k^{\theta})^2+F_{,t} k^{\theta} k^{\theta}_{,t}+\\ 
\frac{1}{2} P_{,tt} (k^{\phi})^2+P_{,t}k^{\phi}k^{\phi}_{,t}+
\Phi_{,tt}k^rk^{\phi}+\Phi_{,t}(k^{\phi}_{,t}k^r+\\
k^{\phi}k^r_{,t})+\Theta_{,tt}k^{\theta}k^r+\Theta_{,t}(k^{\theta}_{,t}k^r+k^{\theta}k^r_{,t})]
\end{split}
\end{equation}
\begin{equation}
\begin{split}
\frac{d}{d\lambda}(k^t_{,r}) =- [k^t_{,t} k^{t}_{,r} + k^t_{,r} k^{r}_{,r} + k^t_{,\theta} k^{\theta}_{,r} + k^t_{,\phi} k^{\phi}_{,r}] -\\
\frac{1}{c^2}[\frac{1}{2} R_{,tr} (k^r)^2+R_{,t} k^r k^r_{,r} + \frac{1}{2} F_{,tr} (k^{\theta})^2+F_{,t} k^{\theta} k^{\theta}_{,r}+\\
\frac{1}{2} P_{,tr} (k^{\phi})^2+
P_{,t}k^{\phi}k^{\phi}_{,r}+\Phi_{,tr}k^rk^{\phi}+\Phi_{,t}(k^{\phi}_{,r}k^r+\\
k^{\phi}k^r_{,r})+\Theta_{,tr}k^{\theta}k^r+
\Theta_{,t}(k^{\theta}_{,r}k^r+k^{\theta}k^r_{,r})]
\end{split}
\end{equation}
\begin{equation}
\begin{split}
\frac{d}{d\lambda}(k^t_{,\theta}) =- [k^t_{,t} k^{t}_{,\theta} + k^t_{,r} k^{r}_{,\theta} + k^t_{,\theta} k^{\theta}_{,\theta} + k^t_{,\phi} k^{\phi}_{,\theta}] -\\
\frac{1}{c^2}[\frac{1}{2} R_{,t\theta} (k^r)^2+R_{,t} k^r k^r_{,\theta} +F_{,t} k^{\theta} k^{\theta}_{,\theta}+ \\
\frac{1}{2} P_{,t\theta} (k^{\phi})^2+P_{,t}k^{\phi}k^{\phi}_{,\theta}+\Phi_{,t\theta}k^rk^{\phi}+\Phi_{,t}(k^{\phi}_{,\theta}k^r+\\
k^{\phi}k^r_{,\theta})+\Theta_{,t\theta}k^{\theta}k^r+\Theta_{,t}(k^{\theta}_{,\theta}k^r+k^{\theta}k^r_{,\theta})]
\end{split}
\end{equation}
\begin{equation}
\begin{split}
\frac{d}{d\lambda}(k^t_{,\phi}) = - [k^t_{,t} k^{t}_{,\phi} + k^t_{,r} k^{r}_{,\phi} + k^t_{,\theta} k^{\theta}_{,\phi} + k^t_{,\phi} k^{\phi}_{,\phi}] -\\
\frac{1}{c^2}[\frac{1}{2} R_{,t\phi} (k^r)^2+R_{,t} k^r k^r_{,\phi} +F_{,t} k^{\theta} k^{\theta}_{,\phi}+ \\ P_{,t}k^{\phi}k^{\phi}_{,\phi}+\Phi_{,t\phi}k^rk^{\phi}+\Phi_{,t}(k^{\phi}_{,\phi}k^r+\\
k^{\phi}k^r_{,\phi})+\Theta_{,t\phi}k^{\theta}k^r+\Theta_{,t}(k^{\theta}_{,\phi}k^r+k^{\theta}k^r_{,\phi})]
\end{split}
\end{equation}
\newline
Equation (\ref{eq:theta}) has been used to obtain $\frac{d}{d\lambda}(k^{\theta}_{,\alpha})$. Inserting $\alpha = t, r, \theta, \phi$ the following four equations for $\frac{d}{d\lambda}(k^{\theta}_{,\alpha})$ are obtained:
\begin{equation}
\begin{split}
\frac{d}{d\lambda} (k^{\theta}_{,t}) = -(k^{\theta}_{,t} k^{t}_{,t} + k^{\theta}_{,r} k^{r}_{,t} + k^{\theta}_{,\theta} k^{\theta}_{,t} + k^{\theta}_{,\phi} k^{\phi}_{,t})-\\
\frac{1}{F}[F_{,t} \dot k^{\theta} + k^{\theta}(F_{,tt} k^{t} +F_{,tr} k^{r}+F_{,t} k^{t}_{,t} + F_{,r} k^{r}_{,t}) +\\ 
\dot F k^{\theta}_{,t} +k^r(\Theta_{,tt} k^{t}+
\Theta_{,rt} k^{r}+\Theta_{,\theta t} k^{\theta}+\Theta_{,\phi t} k^{\phi}+\\
\Theta_{,t} k^{t}_{,t}+\Theta_{,r} k^{r}_{,t}+\Theta_{,\theta} k^{\theta}_{,t}+\Theta_{,\phi} k^{\phi}_{,t}) +\dot \Theta k^r_{,t} +\\
\Theta_{,t} \dot k^r + \Theta (\frac{d}{d\lambda}(k^r_{,t}) + k^r_{,t}k^{t}_{,t}+ k^r_{,r}k^{r}_{,t}+ k^r_{,\theta}k^{\theta}_{,t}+\\ 
k^r_{,\phi}k^{\phi}_{,t}) -  [\frac{1}{2}R_{,\theta t} (k^r)^2 +R_{\theta} k^r k^r_{,t} +\frac{1}{2} P_{,\theta t}(k^{\phi})^2 +\\
P_{,\theta}k^{\phi} k^{\phi}_{,t} +\Theta_{,\theta t}k^{\theta}k^r + \Theta_{,\theta} (k^{\theta}_{,t}k^r + k^{\theta}k^r_{,t}) +\\ \Phi_{,\theta t} k^{\phi}k^r + \Phi_{,\theta} (k^{\phi}_{,t} k^r+k^{\phi}k^r_{,t})]]
\end{split}
\end{equation}
\begin{equation}
\begin{split}
\frac{d}{d\lambda} (k^{\theta}_{,r}) = -(k^{\theta}_{,t} k^{t}_{,r} + k^{\theta}_{,r} k^{r}_{,r} + k^{\theta}_{,\theta} k^{\theta}_{,r} + k^{\theta}_{,\phi} k^{\phi}_{,r})-\\
\frac{1}{F}[F_{,r} \dot k^{\theta} + k^{\theta}(F_{,rt} k^{t} +F_{,rr} k^{r}+ F_{,t} k^{t}_{,r} +\\ 
F_{,r} k^{r}_{,r}) + \dot F k^{\theta}_{,r} +k^r(\Theta_{,tr} k^{t}+
\Theta_{,rr} k^{r}+\Theta_{,\theta r} k^{\theta}+\\
\Theta_{,\phi r} k^{\phi}+\Theta_{,t} k^{t}_{,r}+\Theta_{,r} k^{r}_{,r}+\Theta_{,\theta} k^{\theta}_{,r}+\Theta_{,\phi} k^{\phi}_{,r}) +\\
\dot \Theta k^r_{,r} +\Theta_{,r} \dot k^r + \Theta (\frac{d}{d\lambda}(k^r_{,r}) + k^r_{,t}k^{t}_{,r}+\\ k^r_{,r}k^{r}_{,r}+ k^r_{,\theta}k^{\theta}_{,r}+ k^r_{,\phi}k^{\phi}_{,r}) -  [\frac{1}{2}R_{,\theta r} (k^r)^2 +R_{\theta} k^r k^r_{,r} +\\
\frac{1}{2} P_{,\theta r}(k^{\phi})^2 +P_{,\theta}k^{\phi} k^{\phi}_{,r} +\Theta_{,\theta r}k^{\theta}k^r + \Theta_{,\theta} (k^{\theta}_{,r}k^r + k^{\theta}k^r_{,r}) +\\ 
\Phi_{,\theta r} k^{\phi}k^r + \Phi_{,\theta} (k^{\phi}_{,r} k^r+k^{\phi}k^r_{,r})]]
\end{split}
\end{equation}
\begin{equation}
\begin{split}
\frac{d}{d\lambda} (k^{\theta}_{,\theta}) = -(k^{\theta}_{,t} k^{t}_{,\theta} + k^{\theta}_{,r} k^{r}_{,\theta} + k^{\theta}_{,\theta} k^{\theta}_{,\theta} +\\
k^{\theta}_{,\phi} k^{\phi}_{,\theta})-\frac{1}{F}[k^{\theta}(F_{,t} k^{t}_{,\theta} + F_{,r} k^{r}_{,\theta}) + \dot F k^{\theta}_{,\theta} +\\
k^r(\Theta_{,t\theta} k^{t}+\Theta_{,r\theta} k^{r}+ \Theta_{,\theta\theta} k^{\theta}+\Theta_{,\phi\theta} k^{\phi}+\Theta_{,t} k^{t}_{,\theta}+\\
\Theta_{,r} k^{r}_{,\theta}+\Theta_{,\theta} k^{\theta}_{,\theta}+ \Theta_{,\phi} k^{\phi}_{,\theta}) +\dot \Theta k^r_{,\theta} +
\Theta_{,\theta} \dot k^r +\\ 
\Theta (\frac{d}{d\lambda}(k^r_{,\theta}) + k^r_{,t}k^{t}_{,\theta}+ k^r_{,r}k^{r}_{,\theta}+ k^r_{,\theta}k^{\theta}_{,\theta}+ k^r_{,\phi}k^{\phi}_{,\theta}) -\\
[\frac{1}{2}R_{,\theta\theta} (k^r)^2 +R_{\theta} k^r k^r_{,\theta} +
\frac{1}{2} P_{,\theta\theta}(k^{\phi})^2 +P_{,\theta}k^{\phi} k^{\phi}_{,\theta} +\\
\Theta_{,\theta\theta}k^{\theta}k^r + \Theta_{,\theta} (k^{\theta}_{,\theta}k^r + k^{\theta}k^r_{,\theta}) + \Phi_{,\theta\theta} k^{\phi}k^r +\\ 
\Phi_{,\theta} (k^{\phi}_{,\theta} k^r+k^{\phi}k^r_{,\theta})]]
\end{split}
\end{equation}
\begin{equation}
\begin{split}
\frac{d}{d\lambda} (k^{\theta}_{,\phi}) = -(k^{\theta}_{,t} k^{t}_{,\phi} + k^{\theta}_{,r} k^{r}_{,\phi} + k^{\theta}_{,\theta} k^{\theta}_{,\phi} +\\
k^{\theta}_{,\phi} k^{\phi}_{,\phi})-\frac{1}{F}[k^{\theta}(F_{,t} k^{t}_{,\phi} +F_{,r} k^{r}_{,\phi}) + \dot F k^{\theta}_{,\phi} +\\
k^r(\Theta_{,t\phi} k^{t}+ \Theta_{,r\phi} k^{r}+
\Theta_{,\theta\phi} k^{\theta}+\Theta_{,\phi\phi} k^{\phi}+\Theta_{,t} k^{t}_{,\phi}+\\
\Theta_{,r} k^{r}_{,\phi}+
\Theta_{,\theta} k^{\theta}_{,\phi}+\Theta_{,\phi} k^{\phi}_{,\phi}) +\dot \Theta k^r_{,\phi} +\Theta_{,\phi} \dot k^r +\\ 
\Theta (\frac{d}{d\lambda}(k^r_{,\phi}) + k^r_{,t}k^{t}_{,\phi}+ k^r_{,r}k^{r}_{,\phi}+ k^r_{,\theta}k^{\theta}_{,\phi}+ k^r_{,\phi}k^{\phi}_{,\phi}) -\\
[\frac{1}{2}R_{,\theta\phi} (k^r)^2 +R_{\theta} k^r k^r_{,\phi} + P_{,\theta}k^{\phi} k^{\phi}_{,\phi} +\Theta_{,\theta\phi}k^{\theta}k^r + \\
\Theta_{,\theta} (k^{\theta}_{,\phi}k^r +
k^{\theta}k^r_{,\phi}) + \Phi_{,\theta\phi} k^{\phi}k^r + 
\Phi_{,\theta} (k^{\phi}_{,\phi} k^r+k^{\phi}k^r_{,\phi})]]
\end{split}
\end{equation}
\newline
 Isolating $\frac{d}{d\lambda}(k^{\phi}_{,\alpha})$ in equation (\ref{eq:phi}) and inserting $\alpha = t, r, \phi, \theta$ leads to the following four equations:
\begin{equation}
\begin{split}
\frac{d}{d\lambda}(k^{\phi}_{,t}) = -( k^{\phi}_{,t} k^{t}_{,t} + k^{\phi}_{,r} k^{r}_{,t} + k^{\phi}_{,\theta} k^{\theta}_{,t} + k^{\phi}_{,\phi} k^{\phi}_{,t})-\\
\frac{1}{P}[P_{,t} \dot k^{\phi}+ k^{\phi} (P_{,tt}k^{t} + P_{,rt}k^{r} + P_{,\theta t}k^{\theta} +P_{,\phi t}k^{\phi}+P_{,t}k^{t}_{,t} +\\
P_{,r}k^{r}_{,t} + P_{,\theta}k^{\theta}_{,t} +P_{,\phi}k^{\phi}_{,t}) + \dot P k^{\phi}_{,t} + k^r(\Phi_{,t t}k^{t} +\\
\Phi_{,tr}k^{r} +\Phi_{,t\theta}k^{\theta} +\Phi_{,t\phi}k^{\phi} + \Phi_{,t}k^{t}_{,t} + \Phi_{,r}k^{r}_{,t} +\\ \Phi_{,\theta}k^{\theta}_{,t} + \Phi_{,\phi}k^{\phi}_{,t}) + \dot \Phi k^r_{,t} +
\Phi_{,t} \dot k^r + \Phi(\frac{d}{d\lambda}(k^r_{,t}) +\\ 
k^r_{,t} k^{t}_{,t} + k^r_{,r} k^{r}_{,t} + k^r_{,\theta} k^{\theta}_{,t} + k^r_{,\phi} k^{\phi}_{,t}) -\\
[\frac{1}{2} R_{,\phi t} (k^r)^2+R_{,\phi} k^rk^r_{,t} + 
\Theta_{,\phi t} k^{\theta}k^r + \Theta_{,\phi} (k^{\theta}_{,t}k^r +\\ 
k^{\theta}k^r_{,t}) + \Phi_{,t\phi} k^{\phi}k^r + \Phi_{,\phi}(k^{\phi}_{,t}k^r + k^{\phi}k^r_{,t})]]
\end{split}
\end{equation}

\begin{equation}
\begin{split}
\frac{d}{d\lambda}(k^{\phi}_{,r}) = -( k^{\phi}_{,t} k^{t}_{,r} + k^{\phi}_{,r} k^{r}_{,r} + k^{\phi}_{,\theta} k^{\theta}_{,r} + k^{\phi}_{,\phi} k^{\phi}_{,r})-\\
\frac{1}{P}[P_{,r} \dot k^{\phi}+ k^{\phi} (P_{,tr}k^{t} + P_{,rr}k^{r} + P_{,\theta r}k^{\theta} + P_{,\phi r}k^{\phi}+\\
P_{,t}k^{t}_{,r} +P_{,r}k^{r}_{,r} + P_{,\theta}k^{\theta}_{,r} +P_{,\phi}k^{\phi}_{,r}) + \dot P k^{\phi}_{,r} +\\ 
k^r(\Phi_{,r t}k^{t} +\Phi_{,rr}k^{r} +\Phi_{,r\theta}k^{\theta} +\Phi_{,r\phi}k^{\phi} + \Phi_{,t}k^{t}_{,r} +\\
\Phi_{,r}k^{r}_{,r} + \Phi_{,\theta}k^{\theta}_{,r} + \Phi_{,\phi}k^{\phi}_{,r}) + \dot \Phi k^r_{,r} +
\Phi_{,r} \dot k^r +\\
\Phi(\frac{d}{d\lambda}(k^r_{,r}) + k^r_{,t} k^{t}_{,r} + k^r_{,r} k^{r}_{,r} + k^r_{,\theta} k^{\theta}_{,r} +\\ 
k^r_{,\phi} k^{\phi}_{,r}) - [\frac{1}{2} R_{,\phi r} (k^r)^2+
R_{,\phi} k^rk^r_{,r} + 
\Theta_{,\phi r} k^{\theta}k^r + \Theta_{,\phi} (k^{\theta}_{,r}k^r +\\ 
k^{\theta}k^r_{,r}) + \Phi_{,r\phi} k^{\phi}k^r + \Phi_{,\phi}(k^{\phi}_{,r}k^r + k^{\phi}k^r_{,r})]]
\end{split}
\end{equation}

\begin{equation}
\begin{split}
\frac{d}{d\lambda}(k^{\phi}_{,\theta}) = -( k^{\phi}_{,t} k^{t}_{,\theta} + k^{\phi}_{,r} k^{r}_{,\theta} + k^{\phi}_{,\theta} k^{\theta}_{,\theta} +\\
k^{\phi}_{,\phi} k^{\phi}_{,\theta})-\frac{1}{P}[P_{,\theta} \dot k^{\phi}+ k^{\phi} (P_{,t\theta}k^{t} + P_{,r\theta}k^{r} + P_{,\theta\theta}k^{\theta}+\\ P_{,\phi\theta}k^{\phi}+P_{,t}k^{t}_{,\theta} +P_{,r}k^{r}_{,\theta} + P_{,\theta}k^{\theta}_{,\theta} +P_{,\phi}k^{\phi}_{,\theta}) +\\ \dot P k^{\phi}_{,\theta} + k^r(\Phi_{,\theta t}k^{t} +
\Phi_{,\theta r}k^{r} +\Phi_{,\theta\theta}k^{\theta} +\Phi_{,\theta\phi}k^{\phi} +\\
\Phi_{,t}k^{t}_{,\theta} + \Phi_{,r}k^{r}_{,\theta} + \Phi_{,\theta}k^{\theta}_{,\theta} + \Phi_{,\phi}k^{\phi}_{,\theta}) + \dot \Phi k^r_{,\theta} +\\
\Phi_{,\theta} \dot k^r + \Phi(\frac{d}{d\lambda}(k^r_{,\theta}) + k^r_{,t} k^{t}_{,\theta} + k^r_{,r} k^{r}_{,\theta} + k^r_{,\theta} k^{\theta}_{,\theta} +\\ 
k^r_{,\phi} k^{\phi}_{,\theta}) - [\frac{1}{2} R_{,\phi\theta} (k^r)^2+R_{,\phi} k^rk^r_{,\theta} +
\Theta_{,\phi\theta} k^{\theta}k^r +\\ 
\Theta_{,\phi} (k^{\theta}_{,\theta}k^r + k^{\theta}k^r_{,\theta}) + \Phi_{,\theta\phi} k^{\phi}k^r +\\ 
\Phi_{,\phi}(k^{\phi}_{,\theta}k^r + k^{\phi}k^r_{,\theta})]]
\end{split}
\end{equation}

\begin{equation}
\begin{split}
\frac{d}{d\lambda}(k^{\phi}_{,\phi}) = -( k^{\phi}_{,t} k^{t}_{,\phi} + k^{\phi}_{,r} k^{r}_{,\phi} + k^{\phi}_{,\theta} k^{\theta}_{,\phi} + k^{\phi}_{,\phi} k^{\phi}_{,\phi})-\\
\frac{1}{P}[P_{,\phi} \dot k^{\phi}+ k^{\phi} (P_{,t\phi}k^{t} + P_{,r\phi}k^{r} + P_{,\theta\phi}k^{\theta}+ P_{,\phi\phi}k^{\phi}+\\
P_{,t}k^{t}_{,\phi} +P_{,r}k^{r}_{,\phi} + P_{,\theta}k^{\theta}_{,\phi} +P_{,\phi}k^{\phi}_{,\phi}) + \dot P k^{\phi}_{,\phi} +\\ 
k^r(\Phi_{,\phi t}k^{t} +\Phi_{,\phi r}k^{r} +\Phi_{,\phi\theta}k^{\theta} +\Phi_{,\phi\phi}k^{\phi} + \Phi_{,t}k^{t}_{,\phi} +\\
\Phi_{,r}k^{r}_{,\phi} + \Phi_{,\theta}k^{\theta}_{,\phi} + \Phi_{,\phi}k^{\phi}_{,\phi}) + \dot \Phi k^r_{,\phi} +
\Phi_{,\phi} \dot k^r +\\ 
\Phi(\frac{d}{d\lambda}(k^r_{,\phi}) + k^r_{,t} k^{t}_{,\phi} + k^r_{,r} k^{r}_{,\phi} + k^r_{,\theta} k^{\theta}_{,\phi} + k^r_{,\phi} k^{\phi}_{,\phi}) -\\ 
[\frac{1}{2} R_{,\phi\phi} (k^r)^2 + R_{,\phi} k^rk^r_{,\phi} + 
\Theta_{,\phi\phi} k^{\theta}k^r + \Theta_{,\phi} (k^{\theta}_{,\phi}k^r +\\ 
k^{\theta}k^r_{,\phi}) + \Phi_{,\phi\phi} k^{\phi}k^r + \Phi_{,\phi}(k^{\phi}_{,\phi}k^r + k^{\phi}k^r_{,\phi})]]
\end{split}
\end{equation}
\newline\newline
The Christoffel symbols needed in the ODE for $D_A$ are:
\begin{equation*}
\begin{split}
\Gamma_{\phi t}^{t}= \Gamma_{\theta t}^{t}= \Gamma_{rt}^{t}= \Gamma_{tt}^{t}= 0 \\
\Gamma_{\phi \theta}^{\theta}=-\frac{1}{2F}\frac{\Theta}{R-\frac{\Phi^2}{P}-\frac{\Theta^2}{F}}(\Phi_{,\theta}+\Theta_{,\phi}-\frac{\Phi}{P}P_{,\theta}) \\
\Gamma_{\theta \theta}^{\theta}=-\frac{1}{F}\frac{\Theta}{R-\frac{\Phi^2}{P}-\frac{\Theta^2}{F}}(\Theta_{,\theta}- \frac{1}{2}F_{,r}) \\
\Gamma_{\phi \phi}^{\phi}=-\frac{1}{P}\frac{\Phi}{R-\frac{\Phi^2}{P}-\frac{\Theta^2}{F}}(\Phi_{,\phi}-\frac{1}{2}P_{,r}+\frac{1}{2}\frac{\Theta}{F}P_{,\theta}) \\
\Gamma_{\theta r}^{r}=\frac{1}{2}\frac{1}{R-\frac{\Phi^2}{P}-\frac{\Theta^2}{F}}(R_{,\theta}-\frac{\Phi}{P}\Phi_{,\theta}+\frac{\Phi}{P}\Theta_{,\phi}-\frac{\Theta}{F}F_{,r}) \\
\Gamma_{\phi r}^{r}=\frac{1}{2}\frac{1}{R-\frac{\Phi^2}{P}-\frac{\Theta^2}{F}}(R_{,\phi}-\frac{\Phi}{P}P_{,r}-\frac{\Theta}{F}\Theta_{,\phi}+\frac{\Theta}{F}\Phi_{,\theta}) \\
\Gamma_{tr}^{r}= \frac{1}{2}\frac{1}{R-\frac{\Phi^2}{P}-\frac{\Theta^2}{F}}(R_{,t}-\frac{\Phi}{P}\Phi_{,t}-\frac{\Theta}{F}\Theta_{,t}) \\
\Gamma_{t\theta}^{\theta}= \frac{1}{2F}(F_{,t}-\frac{\Theta}{R-\frac{\Phi^2}{P}-\frac{\Theta^2}{F}}(\Theta_{,t}-\frac{\Theta}{F}F_{,t}))\\
\Gamma_{t\phi}^{\phi}=\frac{1}{2P}(P_{,t}-\frac{\Phi}{R-\frac{\Phi^2}{P}-\frac{\Theta^2}{F}}(\Phi_{,t}-\frac{\Phi}{P}P_{,t})) \\
\Gamma_{rr}^{r}= \frac{1}{2}\frac{1}{R-\frac{\Phi^2}{P}-\frac{\Theta^2}{F}}(\frac{\Phi}{P} R_{,\phi}+\frac{\Theta}{F}R_{,\theta}+\\R_{,r}-2\frac{\Phi}{P}\Phi_{,r}-2\frac{\Theta}{F}\Theta_{,r})\\
\Gamma_{r \theta}^{\theta}=\frac{1}{2F}(F_{,r}-\frac{\Theta}{R-\frac{\Phi^2}{P}-\frac{\Theta^2}{F}}(R_{,\theta}-\\ \frac{\Phi}{P}\Phi_{,\theta}+\frac{\Phi}{P}\Theta_{,\phi} -\frac{\Theta}{F}F_{,r})) \\
\Gamma_{r \phi}^{\phi}=\frac{1}{2P}(P_{,r}-\frac{\Phi}{R-\frac{\Phi^2}{P}-\frac{\Theta^2}{F}}(R_{,\phi}-\\ \frac{\Theta}{F}\Theta_{,\phi}+\frac{\Theta}{F}\Phi_{,\theta}-\frac{\Phi}{P}P_{,r}))\\
\Gamma_{\theta \phi}^{\phi}=\frac{1}{2P}(P_{,\theta}-\frac{\Phi}{R-\frac{\Phi^2}{P}-\frac{\Theta^2}{F}}(\Phi_{,\theta}-\frac{\Phi}{P}P_{,\theta}+\Theta_{,\phi}))
\end{split}
\end{equation*}
\newline\newline
The initial conditions are set using considerations analogous to those made in section V B of \cite{dallas-light}. The result is that $k^t_{,t} = -0.5(\ln R)_{,t}k^t$ initially, while the remaining $k^{\alpha}_{,\beta}$ are zero or determined from the partial derivatives of the null-condition.

\end{document}